\documentclass[10pt, letterpaper]{article}
\usepackage{authblk}
\usepackage{hyperref}
\hypersetup{citecolor=blue}
\hypersetup{
    colorlinks=true,
    linkcolor=blue,
    filecolor=magenta,      
    urlcolor=blue,
    pdfpagemode=FullScreen,
    }
\title{Correlation Functions and Stochastic Feynman Rules for Self-Interacting Scalar Fields}
\author[1]{Moongul Byun\footnote{moongulbyun@hanyang.ac.kr}}
\affil[1]{\small{\textit{Department of Physics, Hanyang University, Seoul, 04763, South Korea}}}
\date{\today}
\usepackage{amsmath, amssymb, braket, tensor, geometry, simpler-wick, graphicx, appendix, blindtext, titlesec, soul, mathtools, enumerate}
\usepackage[english]{babel}
\usepackage{amsthm}
\usepackage{engord}
\usepackage{subcaption}
\usepackage{filecontents}
\usepackage[backend=biber, style=numeric-comp, sorting=none, bibstyle = trad-abbrv]{biblatex}
\newtheorem{remark}{Remark}
\begin{document}

%------------Title----------%
\maketitle
%-------------------------%

%-----------Abstract-----------%
\begin{abstract}
It is well known that perturbative solutions of the Langevin equation can be used to calculate correlation functions in stochastic quantization. However, this work is challenging due to the absence of generalized rules. In this paper, we address this difficulty by studying correlation functions up to certain orders for self-interacting scalar fields. Through the perturbative approach, we establish stochastic Feynman rules applicable to both finite and large fictitious times. Within this process, we introduce a fictitious-time ordering diagram, which serves as a keystone for finding all possible fictitious-time orderings and directly writing down an exact contribution for a given stochastic diagram with its fixed fictitious-time ordering.
\end{abstract}
%-----------------------------%

\newpage

%----------Contents----------%
\tableofcontents
%---------------------------%

%-----------------------Introduction---------------------%
\section{Introduction}
Stochastic quantization is a framework predicting that physical quantities converge to those of quantum field theory (QFT) as the system approaches equilibrium within the reservoir over fictitious-time flows.\cite{parisi1981perturbation, damgaard1987stochastic, namiki1993basic, FLORATOS1983392} A basic idea of stochastic quantization is introducing an additional parameter (i.e. fictitious time or relaxation time $t$) on field operators (i.e. for scalar fields, $\phi(x, t)$) and letting them be described by Langevin equation for a given (Euclidean) action $S_{E}$,\footnote{\label{footnote1}The Langevin equation originally came from the Brownian motion that classically describes a motion of particles subjected to random fluctuations. See Refs.~\cite{uhlenbeck1930theory, lemons1997paul} for more details. Also we observe that the fictitious time coordinate $t$ is non-Euclidean.}
\begin{equation}
\label{eq1}
\frac{\partial}{\partial t} \phi(x, t) = -\frac{\delta S_E}{\delta \phi(x, t)} + \eta(x, t),
\end{equation}
So the main approach to the field $\phi(x, t)$ is that they are governed by the Wiener process, randomized by the Gaussian white noise.\footnote{The Wiener process deals with the continuous-time stochastic process while the Markov process is subjected to the discrete-time stochastic process. For detailed discussions, see Refs.~\cite{Nagasawa, Markov, cinlar1981representation}.} Solving (\ref{eq1}), one can get a perturbative solution of the field\footnote{In another point of view, we can employ partition function (e.g. $Z[\phi(x, t)]$ for scalar fields) and its redefined generator (path-integral formalism) where the corresponding probability distribution can be explained by the Fokker-Planck equation. The Fokker-Planck equation, originally worked in Ref.~\cite{fokker1914mittlere}, is well explained in Refs.~\cite{parisi1981perturbation, damgaard1987stochastic, namiki1993basic, Risken1996} and the functional approach of stochastic quantization can be found in Refs.~\cite{damgaard1987stochastic, gozzi1983functional, Mcclain:1982fb, MENEZES2007617}.} which has a generating source as the random fluctuation, $\eta(x, t)$, and can be represented by stochastic diagrams.\cite{parisi1981perturbation, damgaard1987stochastic, MENEZES2007617, Anzaki2015, HUFFEL1985545, Mcclain:1982fb}

Using the perturbative solutions of the fields, one can find correlation function $\langle \cdots \rangle_{\eta}$ for a given interaction $\mathcal{L}_\text{int}$ and show that the result at equal fictitious times within equilibrium ($t\,\rightarrow\,\infty$) gives the same answer that QFT gives.\cite{parisi1981perturbation, damgaard1987stochastic, namiki1993basic} Furthermore, one can also make a concept of stochastic diagrams analogous to the Feynman diagrams in QFT.\footnote{To see detailed applications and see basic ideas of the stochastic diagram, see Refs.~\cite{parisi1981perturbation, damgaard1987stochastic, MENEZES2007617}.} To obtain the exact answer to the correlation function, we have to consider the following two keywords: fictitious-time orderings and topologically identical stochastic diagrams. The fictitious-time orderings at first appear at connections of the noise fields originating from Wick's theorem for the noise fields.\cite{damgaard1987stochastic} Second, the topologically identical stochastic diagrams should be taken into account since they all contribute to the same order of the correlation function.\cite{damgaard1987stochastic}

Meanwhile, it's challenging to evaluate an exact contribution of a given stochastic diagram since we must find all possible fictitious-time orderings and their each contributions. This necessitate a development of generalized rules for evaluations of contributions in the correlation functions. Although there are several attempts to deal with this problem as suggested in Refs.~\cite{damgaard1987stochastic, HUFFEL1985545, Mcclain:1982fb}, we find that there is a more accessible way to deal with the fictitious-time orderings and to directly write down the contributions for a given stochastic diagram, even covering the contribution at finite fictitious times.

In this paper, we begin by identifying a specific order in the two-point correlation function for the $\phi^3$ theory in stochastic quantization. Initially, we assume that all internal fictitious-time variables are smaller than external fictitious times, i.e., $\forall t > \tau$ or we denote as $\forall t_{\text{ext}} > \tau$. We adopt a symmetry factor, denoted as $S$, analogous to that used in QFT.\cite{peskin2018introduction, Musso:2006pt} This factor originates from Wick's theorem for random fluctuations, $\langle \eta \cdots \eta \rangle_{\eta}$. Subsequently, we shift our focus to the fictitious-time orderings within stochastic diagrams. These fictitious-time orderings appear on connections in the stochastic diagrams and we observe them in our perturbative analysis of the correlation functions. Depending on which fictitious-time ordering is chosen in the calculation, the order in which the integral equation is performed differs. So we get correlation functions for all possible fictitious-time orderings which contribute to non-zero evaluations. By summing the stochastic diagrams for all possible time orderings and topologically identical diagrams, we check that the equilibrium results (at large fictitious times) yield the same answers as those obtained in QFT. Next, we introduce fictitious-time ordering diagrams as useful tools for easily examining all possible fictitious-time orderings for a given stochastic diagram. Based on these fictitious-time ordering diagrams, we propose stochastic Feynman rules at finite fictitious times both $\forall t > \tau$ and $\exists t < \tau$. We observe that all contributions $\exists t < \tau$ go to zero at equal and large fictitious time limits. Then we establish the stochastic Feynman rules at large fictitious times using set notations and justify them. Based on our rules, we find that we need more rigorous generalizations to deal with one-loop contributions of $\phi^4$ theory and unconnected lines where the corresponding fictitious-time variables are independent. We claim that we can directly not only find all possible (trivial and non-trivial) fictitious-time orderings of a given stochastic diagram but also write down the correlation function related to a chosen fictitious-time ordering when we once draw the corresponding fictitious-time ordering diagram through checkmark strategy. We propose ways to support this strategy by considering contractions, unconnected lines, and doubly-mentioned unconnected lines, and finally, we establish the method to read off the fictitious-time ordering diagrams at both finite and large fictitious times. 
%--------------------------------------------------%

%-----------------Correlation Functions for $\phi^3$ Theory in Stochastic Quantization-----------------%
\section{Correlation Functions for $\phi^3$ Theory in Stochastic Quantization}
In this section, we briefly review the stochastic quantization for $\phi^3$ theory and empirically obtain a two-point correlation function under the interaction. This work was proposed in Refs.~\cite{parisi1981perturbation, damgaard1987stochastic}, and we repeat the calculations but more explicitly to cover our theorem in later sections. For $(D + 1)$-dimensional self-interacting $\phi^3$ theory in stochastic quantization, the action in terms of the Euclidean spacetime is given by\footnote{From now on, as we mentioned in footnote \ref{footnote1}, we denote the $(D + 1)$-dimensional spacetime as a meaning of $D$-dimensional Euclidean spacetime in addition to the non-Euclidean fictitious time coordinate.} 
\begin{equation}
\label{eq2}
S_E = \int dt \, d^D x \left[\frac{1}{2}\left( \partial_\mu \phi \right)^2 + \frac{1}{2}m^{2}\phi^2 + \frac{\lambda}{3!} \phi^{3}\right].
\end{equation}
Putting this into the Langevin equation (\ref{eq1}), we find the perturbative solution of the Fourier-transformed scalar field $\phi(k, t)$ as 
\begin{equation}
\label{eq3}
\begin{split}
\phi(k, t) = &\phi_0 (k) \exp[-(k^2 + m^2)t] + \int_{0}^{t}d\tau \exp[-(k^2 + m^2)(t - \tau)]\eta(k, \tau)\\
&-\frac{\lambda}{2!}\int_{0}^{t}d\tau \int \frac{d^D p \, d^D q}{(2\pi)^D} \, \exp[-(k^2 + m^2)(t - \tau)] \, \delta^{(D)} (k - p - q)\\
&\quad \times \left\{\phi_0 (p) \exp[-(p^2 + m^2)\tau] + \int_{0}^{\tau} d\tau' \, \exp[-(p^2 + m^2)(\tau - \tau')]\eta(p, \tau') + \cdots\right\}\\
&\quad \times \left\{\phi_0 (q) \exp[-(q^2 + m^2)\tau] + \int_{0}^{\tau} d\tau'' \, \exp[-(q^2 + m^2)(\tau - \tau'')]\eta(q, \tau'') + \cdots\right\}, 
\end{split}
\end{equation}
where $\phi_{0}(k)\equiv\phi(k, 0)$. Here the fictitious times $\tau$, $\tau'$, $\tau''$, $\cdots$ are internal fictitious time flows for the lines with each corresponding momentum, $k$, $p$, $q$, $\cdots$. The Fourier transformed averaging over the noise fields $\eta$'s is described by the Gaussian distribution so that the correlation functions of the noise fields themselves are\footnote{The probability density for the averaging $\langle\cdots\rangle_\eta$, which is Gaussian, is governed by the Fokker-Planck equation.\cite{parisi1981perturbation, damgaard1987stochastic}} 
\begin{equation}
\label{eq4}
\langle\eta(k, t)\eta(k', t')\rangle_\eta = 2(2\pi)^D \delta^{(D)} (k + k')\delta(t - t'). 
\end{equation}
When we deal with $L$-point correlation functions of $\eta$'s, since the random fluctuations are governed by the Wiener process, we can choose a possible set of (\ref{eq4}) analogous to Wick's theorem in QFT; that is\footnote{This can be referred from concepts of the autocorrelation functions of Gaussian white noises. See Ref.~\cite{marmarelis2004nonlinear} for more details.} 
\begin{equation}
\label{eq5}
\langle \eta (k_1 , t_1) \cdots \eta (k_L , t_L) \rangle_\eta = \sum_{\text{All possible set of }\{i, j\}}^{} \prod_{\{i, j\}}^{} \langle \eta(k_i , t_i) \eta (k_j , t_j)\rangle_{\eta} . 
\end{equation}
Then we can easily find the a single propagator (tree-level two-point correlation function) connecting\footnote{Same method for a general perturbation theory was published in Ref.~\cite{Musso:2006pt}.} $\phi(k_1, t_1)$ and $\phi(k_2, t_2)$ is\footnote{The two possible fictitious-time orderings between $t_1$ and $t_2$ both give non-zero and equal contribution so that we can let equilibrium limits later.} 
\begin{equation}
\label{eq6}
D(k_1 , k_2 , t_1 , t_2) = (2\pi)^D \, \frac{\delta^{(D)} (k_1 + k_2)}{k_{1}^{2} + m^2} \{\exp[ -\vert t_1 - t_2 \vert (k_{1}^{2} + m^2) ] - \exp[-(t_1 + t_2)(k_{1}^{2} + m^2)]\}. 
\end{equation}
When we let the fictitious times be equal and be in equilibrium ($t_1 = t_2 \equiv t\, \rightarrow \, \infty$), (\ref{eq6}) exactly gives the propagator in momentum space in QFT as\footnote{We recognize that such a term as $k^2 + m^2$ inside the exponent for an external $k$-leg is positive definite.} 
\begin{equation}
\label{eq7}
\lim_{t\,\to\,\infty} D(k_1 , k_2 , t) = (2\pi)^D \, \frac{\delta^{(D)} (k_1 + k_2)}{k_{1}^{2} + m^2}.
\end{equation}
%---------------------------------------------------------------------------------------------%

%--------------------Two-Point Correlation Functions for $\phi^3$ Theory in Stochastic Quantization--------------------%
\subsection*{Two-Point Correlation Functions for $\phi^3$ Theory in Stochastic Quantization}
\begin{figure}
\centering
\includegraphics[width = 2.8in]{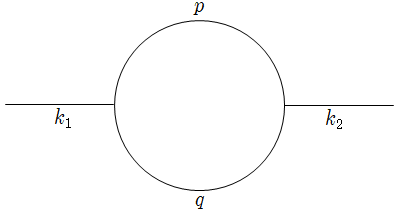}
\caption{\centering{Next-to-leading order (NLO) contribution in the two-point correlation function with $\phi^3$ interaction.}}
\label{Figure 1}
\end{figure}

\begin{figure}
\centering
\begin{subfigure}[b]{0.48\textwidth}
\centering
\includegraphics[width = 2.8in]{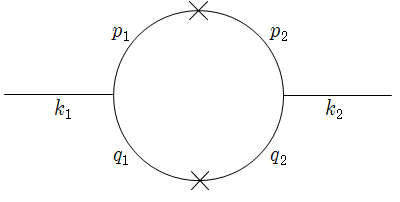}
\caption{\centering{The 1\textsuperscript{st} type of stochastic diagram for NLO in the two-point correlation function with $\phi^3$ interaction.}}
\label{Figure 2a}
\end{subfigure}
\hfill
\begin{subfigure}[b]{0.48\textwidth}
\centering
\includegraphics[width = 2.8in]{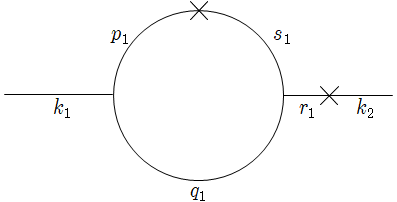}
\caption{\centering{The 2\textsuperscript{nd} type of stochastic diagram for NLO in the two-point correlation function with $\phi^3$ interaction.}}
\label{Figure 2b}
\end{subfigure}
\caption{Two types of stochastic diagram for NLO in two-point correlation function with $\phi^3$ interaction.}
\label{Figure 2}
\end{figure}

We first examine a next-to-leading order (NLO) contribution in the two-point correlation function for $\phi^3$ theory where there are two interactions, suggested in \autoref{Figure 1}. 
First, we consider one of the corresponding stochastic diagrams where the each two branching internal lines are connected, named the 1\textsuperscript{st} type of NLO, $\langle \phi(k_1 , t_1) \phi(k_2 , t_2) \rangle_{\text{NLO}}^{(1)}$ (e.g. \autoref{Figure 2a}) where the superscript $(i)$ means the $i$\textsuperscript{th} type of a given contribution. 
Here the external lines with their momenta $k_1\,(\tau_1 \in [0, t_1])$ and $k_2\,(\tau_2 \in [0, t_2])$ each has two branches with momenta $p_1\,(\tau_{p1} \in [0, \tau_1])$ and $q_1\,(\tau_{q1} \in [0, \tau_1])$, and $p_2\,(\tau_{p2} \in [0, \tau_2])$ and $q_2\,(\tau_{q2} \in [0, \tau_2])$. Therefore, Wick's theorem for the random fluctuations is written by 
\begin{equation}
\label{eq8}
\begin{split}
\langle \eta(p_1 , \tau_{p1}) \eta(p_2 , \tau_{p2}) \eta(q_1 , \tau_{q1}) \eta(q_2 , \tau_{q2}) \rangle_{\eta} =\, & \langle \eta(p_1 , \tau_{p1}) \eta(p_2 , \tau_{p2}) \rangle_{\eta} \langle \eta(q_1 , \tau_{q1}) \eta(q_2 , \tau_{q2}) \rangle_{\eta}\\
& + \, \langle \eta(p_1 , \tau_{p1}) \eta(q_2 , \tau_{q2}) \rangle_{\eta} \langle \eta(p_2 , \tau_{p2}) \eta(q_1 , \tau_{q1}) \rangle_{\eta}\\
& + \, \langle \eta(p_1 , \tau_{p1}) \eta(q_1 , \tau_{q1}) \rangle_{\eta} \langle \eta(p_2 , \tau_{p2}) \eta(q_2 , \tau_{q2}) \rangle_{\eta}. 
\end{split}
\end{equation}
The last term cannot contribute to the correlation function since its corresponding stochastic diagram is composed of two unconnected ones. When we evaluate the first term, i.e., \autoref{Figure 2a}, due to the temporal connections $2\delta(\tau_{p1} - \tau_{p2})$ and $2\delta(\tau_{q1} - \tau_{q2})$, after we conduct the integrals over $\tau_{p2}$ and $\tau_{q2}$, we realize that we should adopt Heaviside step functions, 
\begin{equation}
\label{eq9}
\Theta(\tau_1 - \tau_2) + \Theta(\tau_2 - \tau_1), 
\end{equation}
which represent two possible (internal) fictitious-time orderings between $\tau_1$ and $\tau_2$. These time orderings identically appear in the integral intervals, $\tau_{p1} \in [0, \text{min}(\tau_1, \tau_2)]$ and $\tau_{q1} \in [0, \text{min}(\tau_1, \tau_2)]$. When we choose $\tau_1 > \tau_2$, then the integral range $\tau_2 \in [0, t_2]$ have to be changed to $\tau_2 \in [0, \tau_1]$ for non-zero evaluation. Hence, by gathering the exponential terms of the same $\tau$'s, we find 
\begin{equation}
\label{eq10}
\begin{split}
\langle \phi(k_1 , t_1) \phi(k_2 , t_2) \rangle_{\text{NLO}}^{(1)} = \, &4 (2\pi)^{2D} \left( \frac{\lambda}{2!} \right)^2 \exp[-(k_{1}^{2} + m^2)t_1]\exp[-(k_{2}^{2} + m^2)t_2]\\
& \times \, \int_{0}^{t_1} d\tau_1\int_{0}^{\tau_1} d\tau_2 \int \frac{d^D p_1 \, d^D q_1}{(2\pi)^{2D}} \, \delta^{(D)} (k_1 - p_1 - q_1) \, \delta^{(D)} (k_2 + p_1 + q_1) \\
& \times \, \frac{1}{2}\frac{1}{p_1^2 + m^2}\frac{1}{2}\frac{1}{q_1^2 + m^2}\exp[(k_1^2 - p_1^2 - q_1^2 - m^2)\tau_1]\\
& \times \, \bigg\{\exp[(k_2^2 + p_1^2 + q_1^2 + 3m^2)\tau_2] - \exp[(k_2^2 + p_1^2 - q_1^2 + m^2)\tau_2]\\
& \quad \quad - \exp[(k_2^2 - p_1^2 + q_1^2 + m^2)\tau_2] + \exp[(k_2^2 - p_1^2 - q_1^2 - m^2)\tau_2]\bigg\}\quad(\tau_1 > \tau_2).
\end{split}
\end{equation}
Above, we extract the exponential terms $\exp[-(k_1^2 + m^2)t_1]$ and $\exp[-(k_2^2 + m^2)t_2]$ out of the total integral which are directly related to the external legs. However, when we consider the equal fictitious times $t_1 = t_2 \equiv t$ and let equilibrium ($t \rightarrow \infty$) later, we just need to take care of the first term in the curly bracket for non-zero contribution. Also, when we consider the second term in (\ref{eq8}), we find that this contribution is equal to the first term in (\ref{eq8}). Thus the two contributions in (\ref{eq8}) are symmetric so the overall contribution in $\langle \phi(k_1 , t_1) \phi(k_2 , t_2) \rangle_{\text{NLO}}^{(1)}$ can be obtained by simply multiplying the symmetry factor,\footnote{The original scheme of the symmetry factor for in-in formalism is explained in Ref.~\cite{Musso:2006pt}.} 2, on (\ref{eq10}):\footnote{The two contributions in (\ref{eq8}) share the same fictitious-time orderings (\ref{eq9}). } 
\begin{equation}
\label{eq11}
S = 2 \,, \qquad (\text{for the 1\textsuperscript{st} type of NLO in the two-point correlation functions}). 
\end{equation}
Therefore, by multiplying this symmetry factor, and letting $p_1 \,\rightarrow\, p$ and $q_1 \,\rightarrow\, q$ for simplicity and canceling the common terms, the contribution of $\langle \phi(k_1 , t_1) \phi(k_2 , t_2) \rangle_{\text{NLO}}^{(1)}$ at finite fictitious times under the fictitious-time ordering $(t_{1}, t_{2}) > \tau_1 > \tau_2$ is\footnote{In \hyperlink{subsubsection.3.2.3}{Section 3.2.3} we recast the contributions $\exists t < \tau$ and find they give zero contributions at equal and large fictitious times.} 
\begin{equation}
\label{eq12}
\begin{split}
\langle \phi(k_1 , t_1) \phi(k_2 , t_2) \rangle_{\text{NLO}}^{(1)} = \, &\frac{\lambda^2}{2}\exp[-(k_{1}^{2} + m^2)t_1]\exp[-(k_{2}^{2} + m^2)t_2]\int d^D p \, d^D q \, \frac{\delta^{(D)} (k_1 - p - q)}{p^2 + m^2}\\
& \times \, \frac{\delta^{(D)} (k_2 + p + q)}{q^2 + m^2}\int_{0}^{t_1} d\tau_1 \, \exp[(k_1^2 - p^2 - q^2 - m^2)\tau_1]\\
& \times \, \int_{0}^{\tau_1} d\tau_2 \bigg\{\exp[(k_2^2 + p^2 + q^2 + 3m^2)\tau_2] - \exp[(k_2^2 + p^2 - q^2 + m^2)\tau_2]\\
& \qquad \qquad - \exp[(k_2^2 - p^2 + q^2 + m^2)\tau_2] + \exp[(k_2^2 - p^2 - q^2 - m^2)\tau_2]\bigg\}\quad(\tau_1 > \tau_2).
\end{split}
\end{equation}
At equal and large fictitious times ($t_1 = t_2 \equiv t \,\rightarrow\, \infty$), we find 
\begin{equation}
\label{eq13}
\langle \phi(k_1) \phi(k_2) \rangle_{\text{NLO, Eq}}^{(1)} = \frac{\lambda^2}{2} \int d^D p \, d^D q \,  \frac{\delta^{(D)} (k_1 - p - q)}{p^2 + m^2}\frac{\delta^{(D)} (k_2 + p + q)}{q^2 + m^2}\frac{1}{k_1^2 + k_2^2 + 2m^2}\frac{1}{k_2^2 + p^2 + q^2 + 3m^2}\,(\tau_1 > \tau_2).
\end{equation}
Now we should also consider another time ordering appeared in (\ref{eq9}), $(t_{1}, t_{2}) > \tau_2 > \tau_1$. In this case, the integral interval over $\tau_1$ is changed to $\tau_1 \in [0, \tau_2]$. Also, the symmetry factor $S = 2$ in (\ref{eq11}) identically holds in this case. Therefore, we find\footnote{Or we can derive this contribution for $\tau_1 < \tau_2$ by simply replacing $\tau_1 \,\leftrightarrow\, \tau_2$ from (\ref{eq12}).} 
\begin{equation}
\label{eq14}
\begin{split}
\langle \phi(k_1 , t_1) \phi(k_2 , t_2) \rangle_{\text{NLO}}^{(1)} = \, &\frac{\lambda^2}{2}\exp[-(k_{1}^{2} + m^2)t_1]\exp[-(k_{2}^{2} + m^2)t_2]\int d^D p \, d^D q \, \frac{\delta^{(D)} (k_1 - p - q)}{p^2 + m^2}\\
& \times \, \frac{\delta^{(D)} (k_2 + p + q)}{q^2 + m^2}\int_{0}^{t_2} d\tau_2 \, \exp[(k_2^2 - p^2 - q^2 - m^2)\tau_2]\\
& \times \, \int_{0}^{\tau_2} d\tau_1 \bigg\{\exp[(k_1^2 + p^2 + q^2 + 3m^2)\tau_1] - \exp[(k_1^2 + p^2 - q^2 + m^2)\tau_1]\\
& \qquad \qquad - \exp[(k_1^2 - p^2 + q^2 + m^2)\tau_1] + \exp[(k_1^2 - p^2 - q^2 - m^2)\tau_1]\bigg\}\quad(\tau_1 < \tau_2), 
\end{split}
\end{equation}
and its equilibrium with equal fictitious times gives\footnote{This contribution can be obtained by simply replacing $k_{1}\,\leftrightarrow\,k_{2}$ from (\ref{eq13}).} 
\begin{equation}
\label{eq15}
\langle \phi(k_1) \phi(k_2) \rangle_{\text{NLO, Eq}}^{(1)} = \frac{\lambda}{2} \int d^D p \, d^D q \,  \frac{\delta^{(D)} (k_1 - p - q)}{p^2 + m^2}\frac{\delta^{(D)} (k_2 + p + q)}{q^2 + m^2}\frac{1}{k_1^2 + k_2^2 + 2m^2}\frac{1}{k_1^2 + p^2 + q^2 + 3m^2}\,(\tau_1 < \tau_2).
\end{equation}

We can also imagine another (but topologically identical with \autoref{Figure 2a}) stochastic diagram where one of external legs is branching out into two internal lines and one of them is again branching out into two sub-internal lines, named the 2\textsuperscript{nd} type of NLO, $\langle \phi(k_1 , t_1) \phi(k_2 , t_2) \rangle_{\text{NLO}}^{(2)}$ (e.g. \autoref{Figure 2b}). Here, one of the sub-internal lines is connected to the remaining external leg, while another one is connected to the remaining internal line. We let the external line with momentum $k_1\,(\tau_1 \in [0, t_1])$ have two branching internal lines, $p_1\,(\tau_{p1} \in [0, \tau_1])$ and $q_1\,(\tau_{q1} \in [0, \tau_1])$, and the last line also have two branches, $r_1\,(\tau_{r1} \in [0, \tau_{q1}])$ and $s_1\,(\tau_{s1} \in [0, \tau_{q1}])$. The corresponding Wick's theorem for the random fluctuations is, 
\begin{equation}
\label{eq16}
\begin{split}
\langle\eta(p_1 , \tau_{p1})\eta(r_1 , \tau_{r1})\eta(s_1 , \tau_{s1})\eta(k_2 , \tau_2)\rangle_{\eta} =& \langle\eta(p_1 , \tau_{p1})\eta(s_1 , \tau_{s1})\rangle_{\eta}\langle\eta(r_1 , \tau_{r1})\eta(k_2 , \tau_2)\rangle_{\eta}\\
& + \langle\eta(p_1 , \tau_{p1})\eta(r_1 , \tau_{r1})\rangle_{\eta}\langle\eta(s_1 , \tau_{s1})\eta(k_2 , \tau_2)\rangle_{\eta}\\
& + \langle\eta(p_1 , \tau_{p1})\eta(k_2 , \tau_2)\rangle_{\eta}\langle\eta(r_1 , \tau_{r1})\eta(s_1 , \tau_{s1})\rangle_{\eta}. 
\end{split}
\end{equation}
\autoref{Figure 2b} is the case of the first term in (\ref{eq16}). We note that the second term in (\ref{eq16}) just represents the topologically identical diagram with that for the first term in (\ref{eq16}), and since the third term in (\ref{eq16}) cannot directly contribute to the correlation function, we find that the symmetry factor for this contribution is also 2:
\begin{equation}
\label{eq17}
S = 2 \,, \qquad (\text{for the 2\textsuperscript{nd} type of NLO in the two-point correlation functions}). 
\end{equation}
From \autoref{Figure 2b}, we find that the fictitious-time orderings appear on the two connections as two trivial orderings in the form of
\begin{equation}
\label{eq18}
\Theta(t_2 - \tau_{q1})\Theta(\tau_1 - \tau_{q1}). 
\end{equation}
After applying these trivial time orderings, the temporal integral intervals over $\tau_{p1}$ and $\tau_2$ are changed as $\tau_{p1} \in [0, \text{min}(\tau_1 , \tau_{q1}) = \tau_{q1}]$ and $\tau_2 \in [0, \text{min}(t_2 , \tau_{q1}) = \tau_{q1}]$. Following those fictitious-time orderings and multiplying the symmetry factor, we find the NLO contribution for the second type is (also substituting $p_1 \,\rightarrow\, p$ and $q_1 \,\rightarrow\, q$ for simplicity)
\begin{equation}
\label{eq19}
\begin{aligned}
\langle \phi(k_1 , t_1) \phi(k_2 , t_2) \rangle_{\text{NLO}}^{(2)} = \, &\frac{\lambda^2}{2}\exp[-(k_1^2 + m^2)t_1]\exp[-(k_2^2 + m^2)t_2]\int d^D p \, d^D q \, \frac{\delta^{(D)} (k_1 - p - q)}{p^2 + m^2}\\
&\times \frac{\delta^{(D)} (k_2 + p + q)}{k_2^2 + m^2}\int_{0}^{t_1}d\tau_1 \exp[(k_1^2 - p^2 - q^2 - m^2)\tau_1]\\
&\times \int_{0}^{\tau_1}d\tau_{q1} \bigg\{\exp[(k_2^2 + p^2 + q^2 + 3m^2)\tau_{q1}] - \exp[(k_2^2 - p^2 + q^2 + m^2)\tau_{q1}]\\
& \qquad \qquad - \exp[(k_2^2 + p^2 - q^2 + m^2)\tau_{q1}] + \exp[(k_2^2 - p^2 - q^2 - m^2)\tau_{q1}]\bigg\}.
\end{aligned}
\end{equation}
We find that the first term inside the curly bracket can only contribute to the non-zero correlation function when we impose equal and large fictitious times later. Therefore, letting $t_1 = t_2 \equiv t \,\rightarrow\, \infty$, we find 
\begin{equation}
\label{eq20}
\langle \phi(k_1) \phi(k_2) \rangle_{\text{NLO, Eq}}^{(2)} = \frac{\lambda^2}{2} \int d^D p \, d^D q \, \frac{\delta^{(D)} (k_1 - p - q)}{p^2 + m^2}\frac{\delta^{(D)} (k_2 + p + q)}{k_2^2 + m^2}\frac{1}{k_1^2 + k_2^2 + 2m^2}\frac{1}{k_2^2 + p^2 + q^2 + 3m^2}.
\end{equation}

\begin{figure}
\centering
\includegraphics[width = 6.5in]{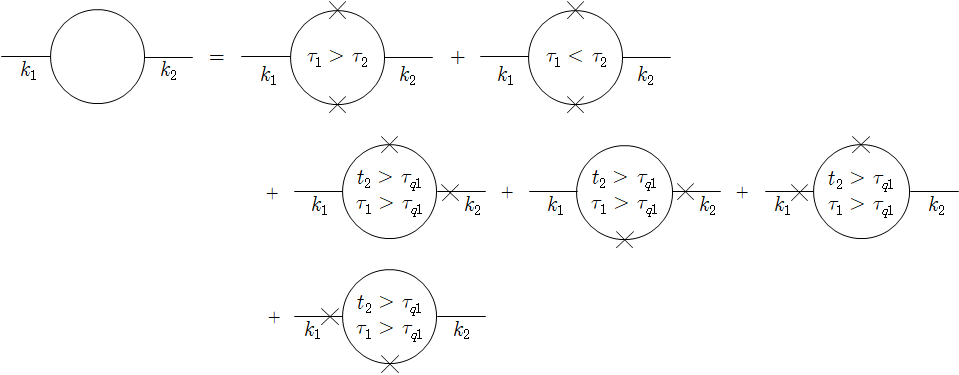}
\caption{\centering{All contributing stochastic diagrams of NLO in the two-point correlation functions for $\phi^3$ interaction.}}
\label{Figure 3}
\end{figure}

Now we have two kinds of NLO in the two-point correlation function. When we expand the correlation function at large fictitious times $\langle\phi(k_1)\phi(k_2)\rangle_{\eta}$ perturbatively, we need to sum all of the types of each fictitious-time ordered correlation function as \autoref{Figure 3}.
We already calculated the first, second, and third diagrams in RHS in \autoref{Figure 3}. The fourth diagram can be found by simply exchanging $p\,\leftrightarrow\,q$ from the third one. The fifth diagram can be found by exchanging $k_1\,\leftrightarrow\,k_2$ from the third one, and the sixth diagram can be evaluated by exchanging $p\,\leftrightarrow\,q$ from the fifth one. Then we finally get LHS in \autoref{Figure 3} as  
\begin{equation}
\label{eq21}
\langle\phi(k_1)\phi(k_2)\rangle_{\text{NLO, Eq}} = \frac{\lambda^2}{2} \frac{1}{k_1^2 + m^2} \frac{1}{k_2^2 + m^2} \int d^D p \int d^D q \, \frac{\delta^{(D)} (k_1 + p + q)}{p^2 + m^2} \frac{\delta^{(D)} (k_2 - p - q)}{q^2 + m^2},
\end{equation}
which is the same as NLO in the two-point correlation function for $\phi^3$ theory in the Euclidean QFT. We recognize that to find out some correlation function specified with a stochastic diagram, we should consider fictitious-time orderings and figure out all possible topologically identical stochastic diagrams within the perturbation. 
%--------------------------------------------------------------------------------%

%--------------------------------Establishment of Stochastic Feynman Rules--------------------------------%
\section{Establishment of Stochastic Feynman Rules}
In this section, we establish stochastic Feynman rules that can effectively explain stochastic diagrams and correlation functions. We empirically find some rules from calculating correlation functions for $\phi^3$ theory (e.g. processes that we did in \hyperlink{section.2}{Section 2}). To do this work, we suggest a diagrammatic way to determine all possible fictitious-time orderings (they give different contributions within the correlation function) and to write down propagators for a given stochastic diagram at both finite and large fictitious times, i.e., \textbf{fictitious-time ordering diagram}. We also suggest methods how to read off the fictitious-time ordering diagram. 
%-----------------------------------------------------------------------------------------------------%

%-----------------------------------Fictitious-Time Ordering Diagrams-----------------------------------%
\subsection{Fictitious-Time Ordering Diagrams}
In this section, we suggest the simplest way to deal with the fictitious-time orderings by introducing fictitious-time ordering diagrams. Within the process of conducting calculations in \hyperlink{section.2}{Section 2}, we discussed the fictitious-time orderings at each connection (some of them gave non-zero contributions while others give zero contributions). However, among the contributing fictitious-time orderings, there are non-trivial undetermined orderings that we have to take a consideration while others are trivial. To determine which the orderings are what we have to deal with, we have to consider their temporal lengths and connections between the stochastic diagrams. As an example, we consider a specific stochastic diagram; that is, \autoref{Figure 2}, previously discussed. 
%--------------------------------------------------------------------------------------------------%

%Fictitious-Time Ordering Diagram for The 1\textsuperscript{st}-Type of NLO in Two-Point Correlation function with $\phi^{3}$-Theory%
\subsubsection*{Fictitious-Time Ordering Diagram for The 1\textsuperscript{st}-Type of NLO in Two-Point Correlation function with $\phi^{3}$-Theory}
\begin{figure}
\centering
\includegraphics[width = 4.7in]{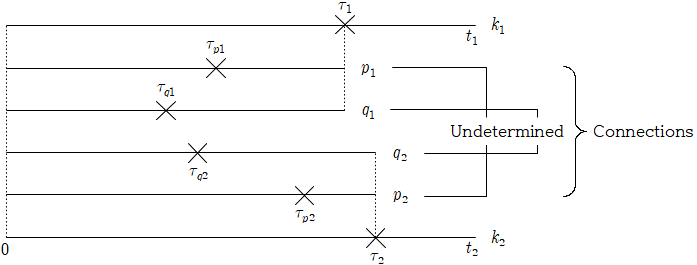}
\caption{\centering{Fictitious-time ordering diagram for the connections occurred in \autoref{Figure 2a}.}}
\label{Figure 4}
\end{figure}

For instance, we can draw a fictitious-time ordering diagram that appeared on the contractions in \autoref{Figure 2a} as \autoref{Figure 4}. 
In this figure, the fictitious times whose orderings between them are undetermined are $\tau_1$ and $\tau_2$. Then we have to consider two possible fictitious-time orderings $\Theta(\tau_1 - \tau_2) + \Theta(\tau_2 - \tau_1)$ as we already discussed in (\ref{eq9}). This can be observed in \autoref{Figure 4}. When we imagine \autoref{Figure 2a} which has fictitious time scales embedded in the propagators, we can directly correspond the lengths of the internal propagators to the evolution of the internal fictitious times. When we increase the length of the propagators with momenta $p_1$ and $q_1$ then the connections will occur at the right side of the internal loop which corresponds to the case where $\tau_1 > \tau_2$ and vice versa. Because the two cases do not disturb the simultaneous connections, the undetermined fictitious-time orderings appear for this reason. They both contribute to the two-point correlation function so we have to sum both fictitious-time ordered stochastic diagrams as we showed in \autoref{Figure 3}.
%----------------------------------------------------------------------------------------------------------------%

%-----------------------------Stochastic Feynman Rules at Finite Fictitious Times-----------------------------%
\subsection{Stochastic Feynman Rules at Finite Fictitious Times}
In this section, we make a foundation of the stochastic Feynman rules at finite fictitious times $\forall t > \tau$ and $\exists t < \tau$, not in equilibrium. To do this, we first suggest the simplest way to write down propagator terms of a specific contribution in the correlation function we are interested in, only from the fictitious-time ordering diagram. We then apply this scheme to the two-point correlation function in \autoref{Figure 2a} and verify its justification. From now on, for notational simplicity, we use symbols $k^2 + m^2$ and $k^2$ interchangeably.\cite{damgaard1987stochastic}
%-----------------------------------------------------------------------------------------------------%

%----------------------Strategies to Write Down Propagators at Finite Fictitious Times $\forall t > \tau$----------------------%
\subsubsection{Strategies to Write Down Propagators at Finite Fictitious Times $\forall t > \tau$}
When we expand the correlation function in the form of integrals, as we can refer from \hyperlink{section.2}{Section 2}, we observe that the terms where the connections occur can be integrated first independently of other integrals. To be specific, when we imagine a case where two propagators with momenta $k_1$ and $k_2$ are connected, whose fictitious-time intervals are each $\tau_1 \in [0, t_1]$ and $\tau_2 \in [0, t_2]$, then we can first apply the momentum conservation law $(2\pi)^D \delta^{(D)}(k_1 + k_2)$ and their temporal connection, $2\delta(\tau_1 - \tau_2)$. After we impose notational simplicity $k_1^2 = k_2^2 \equiv k^2$ and $\tau_1 = \tau_2 \equiv \tau$, the connection $\forall t > \tau$ is described by 
\begin{equation}
\label{eq22}
\text{(Connection between $k_1$ and $k_2$)}\,:\,\int_0^{\text{min}(t_{1}, t_{2})}d\tau \, \exp[-(k^2 + m^2)(t_1 + t_2 - 2\tau)],
\end{equation}
which can be evaluated independently of the other integrals. We see that fictitious-time orderings $\Theta(t_1 - t_2) + \Theta(t_2 - t_1)$ we have to choose appear on this connection. When we choose $t_1 > t_2$, then this yields 
\begin{equation}
\label{eq23}
\text{(Connection between $k_1$ and $k_2$)}\,:\,\frac{1}{2}\frac{1}{k^2 + m^2} \{\exp[-(k^2 + m^2)(t_1 - t_2)] - \exp[-(k^2 + m^2)(t_1 + t_2)]\}.
\end{equation}
The factor $1/2$ will be canceled with a factor $2$ ahead of the temporal connection. So we first conclude that
\begin{enumerate}
\item After applying momentum conservation laws, for propagator with momentum $k$ where stochastic connection occurs, write down 
\begin{equation}
\label{eq24}
\frac{1}{k^2 + m^2}.
\end{equation}
\end{enumerate}
This term can be directly extracted from the fictitious-time ordering diagrams. For instance, from \autoref{Figure 4}, after applying the momentum conservation laws, $p_{1}^{2} = p_{2}^{2} \equiv p^{2}$ and $q_{1}^{2} = q_{2}^{2} \equiv q^{2}$, we can directly write down propagators representing the connected fields as 
\begin{equation}
\label{eq25}
\langle\phi(k_1 , t_1)\phi(k_2 , t_2)\rangle_{\text{NLO}}^{(1)}\,\sim\,\frac{1}{p^2}\frac{1}{q^2}.
\end{equation}

Now there are remaining two lines with momenta $k_1$ and $k_2$ which have no connection. To impose the corresponding propagators, we must first consider the possible fictitious-time orderings of the given stochastic diagram. These orderings can be found by observing the trivial or non-trivial fictitious-time orderings among the connections as we discussed in \hyperlink{subsection.3.1}{Section 3.1}. We have to take integrals over those fictitious-time variables. After picking one of the possible fictitious-time orderings, we can make formulas for the terms of the propagators with no connection, starting from the minimum fictitious-time variable. For the unconnected propagator with its momentum $k$ which has the minimum fictitious-time variable $\tau_{\text{min}} \in [0 , \tau_{\text{max}}]$,\footnote{We adopt the notation $\tau_{\text{max}}$ since, even though evolution of the momentum $k$ runs $\tau_{\text{min}} \in [0 , t]$, once the time ordering $\tau_{\text{min}} < \tau_{\text{max}}$ is determined, the fictitious time interval is changed to $\tau_{\text{min}} \in [0 , \tau_{\text{max}}]$.} the corresponding term within the correlation function is described in the form of 
\begin{equation}
\label{eq26}
\text{(Propagator with momentum $k$ with no connection)}\,:\,\int_{0}^{\tau_{\text{max}}}d\tau_{\text{min}} \, \exp[-(k^2 + m^2)(\tau_{\text{max}} - \tau_{\text{min}})].
\end{equation}
Meanwhile, this $k$-line will branch out into two $k_{\text{branched}}$-lines in $\phi^3$ theory. Since the $k$-line has the smallest fictitious-time variable, then the both branched lines would have connecting spots at the ends of the branches (if not, either one of the branches has no connection, or neither them has no connection, then the fictitious-time orderings will be messed up). Furthermore, in (\ref{eq23}), which is the terms for the branched lines which have the connections, the minimum fictitious-time variable will be appeared in $t_2$, saying $t_2 = \tau_{\text{min}}$. Thus the corresponding terms $\pm\exp(\pm k_{\text{branched}}^2 t_2)$ in (\ref{eq23}) will be gathered with the term $\exp(k^{2} \tau_\text{min})$ in (\ref{eq26}). Then we have to integrate over the variable $t_2 = \tau_{\text{min}}$ in (\ref{eq26}) first, and then, integrate over $\text{max}(t_1 , t_2) = t_1 = \tau_{\text{max}}$.\footnote{The next smallest fictitious-time variable $\tau_{\text{max}}$ stands for another unconnected propagator where the connected line is branched out. We recognize that all connected lines are branched out from their each unconnected propagator.} In other words, we have to consider a term where one $\exp(k^{2} \tau_{\text{min}})$ term for the unconnected line in (\ref{eq26}) and two (\ref{eq23}) terms for the connected lines, each representing one of the branched lines from $k$, are multiplied. By noticing the sign ahead of the each ``$\exp$" and $t_2 = \tau_{\text{min}}$ inside the exponents in (\ref{eq23}), the terms in the correlation functions for the unconnected line with its momentum $k$, which has its branching lines with momenta $(p , q) \in k_{\text{branched}}$ and contains the minimum fictitious-time variable $\tau_{\text{min}}$, can be written as 
\begin{enumerate}
\item[{2. }] Starting from the minimum fictitious-time variable $\tau_{\text{min}} < \tau_{\text{max}}$, for an unconnected propagator $k$ branching to propagators with momenta $p$ and $q$, impose 
\begin{equation}
\label{eq27}
\int_{0}^{\tau_{\text{max}}}d\tau_{\text{min}}\left\{\sum_{\xi_{p},\,\xi_{q}\,=\,\pm1} \xi_{p}\,\xi_{q}\exp\left[(k^{2} + \xi_{p}\,p^{2} + \xi_{q}\,q^{2})\tau_{\text{min}}\right]\right\}.
\end{equation}
\end{enumerate}

To comprehend the above scheme for unconnected lines in a diagrammatic way, we apply this to our example, \autoref{Figure 2a} together with \autoref{Figure 4}. When we choose $(t_1, t_2) > \tau_1 > \tau_2$ in (\ref{eq9}), among the unconnected lines $k_1$ and $k_2$, the later one has the minimum fictitious-time variable $\tau_2$. After applying the momentum conservation laws, because the propagator with $k_2$ branches out to the propagators with momenta $p$ and $q$, we impose 
\begin{equation}
\label{eq28}
\int_{0}^{\tau_1} d\tau_2\left\{\exp\left[(k_2^2 + p^2 + q^2)\tau_2\right] - \exp\left[(k_2^2 - p^2 + q^2)\tau_2\right] - \exp\left[(k_2^2 + p^2 - q^2)\tau_2\right] + \exp\left[(k_2^2 - p^2 - q^2)\tau_2\right]\right\}.
\end{equation}
It's convenient to put check marks on the time ordering diagram for the mentioned momenta in (\ref{eq28}), $k_2$, $p_2$, and $q_2$. We notice that the momenta $p_1$ and $q_1$ are checked automatically when we put check marks on $p_2$ and $q_2$ due to the momentum conservation laws following the connections. 

To find terms for the other unconnected propagators, we should conduct the integrals in order of smaller fictitious time; that means, now we should integrate the equations over $t_1 = \tau_{\text{max}}$ in (\ref{eq23}) or (\ref{eq27}) which is the second smallest fictitious-time variable. The overall forms of the propagators follow (\ref{eq27}) identically. However, if the unconnected propagator branches out to the already mentioned momenta in (\ref{eq27}), then the signs ahead of the doubly-mentioned momenta inside the exponent are equally fixed as a minus sign. The reason can be found in (\ref{eq23}) and (\ref{eq26}). If the line branched out from the unconnected line is what we already mentioned, then there are two possible cases; the branched line has a connection or not. If the branched line has the connection which was represented by (\ref{eq23}), the term for the next minimum fictitious time $t_1 = \tau_{\text{max}}$ is equally imposed in the form of $\exp(-k^2 \tau_{\text{max}})$ and this term can be extracted out of the previously conducted integral over $\tau_{\text{min}}$, while the remaining terms in (\ref{eq23}) were already used in (\ref{eq27}). Therefore, the doubly-mentioned momentum within the integral over the next minimum fictitious time has a fixed ``$-$" sign ahead of its squared term. On the other hand, if the doubly-mentioned branched propagator has no connection, then this indicates that the branched propagator was already evaluated following by (\ref{eq26}). In this integral, we remind that the term $\exp(k^{2} \tau_{\text{min}})$ was used when we conduct the integral for the unconnected propagator which has the smallest fictitious-time variable, $\tau_{\text{min}}$. The remaining term $\exp(-k^{2} \tau_{\text{max}})$ now can be melted into the integral for the second smallest fictitious-time variable, $\tau_{\text{max}}$. So, also for the case where the branched line is unconnected one, the doubly-mentioned momentum squared term has a fixed ``$-$" sign ahead of it. 

By consolidating the above two cases, we conclude that the other unconnected propagators are equally described by (\ref{eq27}) but fixed ``$-$" signs ahead of the doubly-mentioned momentum terms: 
\begin{enumerate}
\item[{3. }] Starting from the minimum fictitious-time variable $\tau_{\text{min}} < \cdots < \tau_{i} < \tau_{j} < \cdots$, for an unconnected propagator $k_{i}$ branching to propagators with momenta $p_{i}$ and $q_{i}$, impose 
\begin{equation}
\label{eq29}
\int_{0}^{\tau_{j}}d\tau_{i}\left\{\sum_{\xi_{p},\,\xi_{q}\,=\,\pm1} \xi_{p}\,\xi_{q} \exp\left[\left(k_{i}^{2} + \xi_{p}\,p_{i}^{2} + \xi_{q}\,q_{i}^{2}\right)\tau_{i}\right]\right\}.
\end{equation}
\textbf{Note} : For doubly-mentioned momenta, the corresponding signs are fixed as $\xi = -1$.
\end{enumerate}
Here, the sign in front of each exponential is determined by the number of minus signs inside the exponent, not counting the sign ahead of the doubly-mentioned momenta; this is because the signs ahead of exponential terms for the already mentioned momenta can identically be extracted with ``$+$" signs in both (\ref{eq23}) and (\ref{eq26}) as $\exp(-k^2 t_1)$. This is a generalization of (\ref{eq27}). So, when momenta are doubly marked on the fictitious-time ordering diagram, we impose minus signs on the corresponding momentum-squared terms. We denote the putting check mark method introduced in this section as \textbf{checkmark strategy}.

We apply this strategy to \autoref{Figure 2a} and \autoref{Figure 4}. Following the previous fictitious-time ordering, after we explore up to (\ref{eq28}), the remaining unconnected propagator is that with its momentum $k_1$ branching out to the $p$- and $q$-line. However, since the momenta $p$ and $q$ were already mentioned in (\ref{eq28}), that means we already once put check marks on each $p_1$ ($ = -p_{2}$) and $q_1$ ($ = -q_{2}$) on \autoref{Figure 4}, the signs ahead of $p^2$ and $q^2$ inside the exponent for $k_1$ are both fixed as a ``$-$" sign. Then there are no more possible pairs of $\pm$ ahead of the exponent and squared momenta so we have to write down the term for the propagator with momentum $k_1$ as (since $k_1$ has its fictitious-time range $\tau_1 \in [0, t_1]$)
\begin{equation}
\label{eq30}
\int_{0}^{t_1}d\tau_1 \, \exp[(k_1^2 - p^2 - q^2)\tau_1].
\end{equation}
The integral forms (\ref{eq28}) and (\ref{eq30}) can exactly be checked in (\ref{eq12}).
%-------------------------------------------------------------------------------------------------------------%

%--------------------------Stochastic Feynman Rules at Finite Fictitious Times $\forall t > \tau$--------------------------%
\subsubsection{Stochastic Feynman Rules at Finite Fictitious Times $\forall t > \tau$}
%---------------------------------------------------------------------------------------------------------------%

%-------------------Stochastic Feynman Rules at Finite Fictitious Times $\forall t > \tau$ in Integral Forms-------------------%
\subsubsection*{Stochastic Feynman Rules at Finite Fictitious Times $\forall t > \tau$ in Integral Forms}
Following the checkmark strategy in the previous section, we suggest stochastic Feynman rules at finite fictitious times $\forall t > \tau$ for the Euclidean field theory as below:
\begin{enumerate}
\item For $(D + 1)$-dimensional spacetime, impose momentum conservation laws, 
\begin{equation}
\label{eq31}
\delta^{(D)} (k + \Sigma_{\text{branched}}\,p)\quad(\text{at each vertex}),\qquad\delta^{(D)} (k_1 + k_2)\quad(\text{at each connection}).
\end{equation}
\item For $\phi^n$ theory, at each vertex, attach 
\begin{equation}
\label{eq32}
-\frac{\lambda}{(n - 1)!}.
\end{equation}
\item For each external line $k_{\text{ext}}$ whose fictitious-time range is $\tau_{\text{ext}}\in[0, t_{\text{ext}}]$, multiply the corresponding factor
\begin{equation}
\label{eq33}
\exp(-k_{\text{ext}}^2 t_{\text{ext}}).
\end{equation}
\item After applying the momentum conservation laws, at each connection with momentum $k_{\text{c}}$, write down 
\begin{equation}
\label{eq34}
\frac{1}{k_{\text{c}}^2}.
\end{equation}
\item For $\phi^{n}$ theory, among the unconnected propagators $k_1\,(0 \leq \tau_1 \leq t_1)$, $\cdots$, $k_N\,(0 \leq \tau_N \leq t_N)$ with a given fictitious-time ordering $\forall t_{\text{ext}} > \tau_1 > \cdots > \tau_N > 0$, pick the momentum $k_N$ which has the smallest fictitious-time variable $\tau_{\text{min}} = \tau_N$ as a starting point and impose 
\begin{equation}
\label{eq35}
\int_{0}^{\tau_{N - 1}} d\tau_N \left\{\sum_{\xi_{N, 1},\,\cdots\,,\,\xi_{N, n - 1}\,=\,\pm1}\left(\prod_{j = 1}^{n - 1}\xi_{N,\,j}\right)\exp\left[\left(k_N^2 + \sum_{i = 1}^{n - 1}\,\left(\xi_{N, i}\,p_{N, i}^{2}\right)\right)\tau_N\right]\right\}
\end{equation}
where $p_{N, 1}, \cdots, p_{N, n - 1}$ are the branched internal lines from $k_N$.
\item For the other unconnected propagators, in order of which has a smaller fictitious-time variable, impose 
\begin{equation}
\label{eq36}
\int_{0}^{\tau_{i - 1}} d\tau_i \left\{\sum_{\xi_{i, 1},\,\cdots\,,\,\xi_{i, m}\,=\,\pm1}\left(\prod_{l = 1}^{m}\xi_{i,\,l}\right)\exp\left[\left(k_i^2 + \sum_{j = 1}^{m}\,\left(\xi_{i, j}\,p_{i, j}^{2}\right) - \sum_{k = m + 1}^{n - 1}\,\left(q_{i, k}^2\right)\right)\tau_{i}\right]\right\}
\end{equation}
where $p_{i, 1}, \cdots, p_{i, m}$ are not doubly-mentioned internal lines while $q_{i, m + 1}, \cdots, q_{i, n - 1}$ are doubly-mentioned internal lines, both branched from $k_{i}$.
\textbf{Note} : For the lastly remaining external leg $k_1$, the integral range should be $\tau_1 \in [0, t_1]$.
\item For each undetermined momentum $p$, take an integral, 
\begin{equation}
\label{eq37}
\int d^{D} p.
\end{equation}
\item For $\phi^n$ theory, multiply 
\begin{equation}
\label{eq38}
(2\pi)^{D \, \times \, \text{[(number of all connections) - $n$(number of all vertices)]}}.
\end{equation} 
\item Multiply by a symmetry factor, 
\begin{equation}
\label{eq39}
S.
\end{equation}
\end{enumerate}
We notice that the other number factors will be canceled by the constants that came from the conducting integrals analogous to a description in Ref.~\cite{HUFFEL1985545}
\subsubsection*{Stochastic Feynman Rules at Finite Fictitious Times $\forall t > \tau$ in Fully Expanded Forms}
Following the previous rules, we expand the integrals for unconnected lines $\forall t > \tau$. From the 5\textsuperscript{th} rule, we first expand it which stands for the minimum fictitious-time variable $\tau_{N}$ so we get total $2^{\,n - 1}$ terms as 
\begin{equation}
\begin{aligned}
\label{eq40}
&\exp\left[\left(k_{N}^{2} + p_{N, 1}^{2} + \cdots + p_{N, n - 1}^{2}\right)\tau_{N - 1}\right] - 1\,,\quad-\left\{\exp\left[\left(k_{N}^{2} - p_{N, 1}^{2} + \cdots + p_{N, n - 1}^{2}\right)\tau_{N - 1}\right] - 1\right\},\\
&\qquad\qquad\qquad\qquad\qquad\qquad\qquad\qquad\qquad\cdots\,,\,(-1)^{n - 1}\left\{\exp\left[\left(k_{N}^{2} - p_{N, 1}^{2} - \cdots - p_{N, n - 1}^{2}\right)\tau_{N - 1}\right] - 1\right\}.
\end{aligned}
\end{equation}
The each terms give their own propagator terms with corresponding denominators $k_N^2 + \xi_{N, 1}\,p_{N, 1}^{2} + \cdots + \xi_{N, n - 1}\,p_{N, n - 1}^{2}$. When we suppose that the internal line $p_{N, 1}$ is doubly-mentioned momentum between $k_N$ and $k_{N - 1}$, then we get total $2^{\,n - 1}$ terms for $k_{N - 1}$ in similar forms of (\ref{eq40}). Meanwhile the original integral form for $k_{N - 1}$-line has its exponent as $k_{N - 1}^2 - p_{N, 1}^{2} \pm \cdots \pm p_{N, n - 1}^{2}$ so we find that it's useful to map this exponent onto each $2^{\,n - 1}$ exponential terms and equal number of $-1$ terms in (\ref{eq40}). After the mapping, the summed squared-momenta will be extracted out as denominators of propagators when we conduct $\tau_{N - 2}$-integrals and additional $-1$ term appears within the each term due to the initial temporal boundaries. We repeat this up to $k_1$ which gives propagators with corresponding final exponential terms and $-1$ where the each denominator of the propagator and corresponding exponent are identical. Then we just need to sum all mapped terms.\footnote{A detailed example is suggested in \hyperlink{subsection.A.3}{A.3}.} We notice that the each exponential terms and $-1$ terms cannot be mapped onto $-1$.
%-------------------------------------------------------------------------------------------------------------%

%------------------Strategies to Write Down Propagators at Finite Fictitious Times $\exists t < \tau$------------------%
\subsubsection{Strategies to Write Down Propagators at Finite Fictitious Times $\exists t < \tau$}
So far we discussed the special cases where every given (or fixed) fictitious-time ranges $t$'s are larger than every internal fictitious-time variables $\tau$'s. In this section, we rediscover our checkmark strategy for the cases where some (or every) $t$'s are smaller than some (or every) $\tau$'s, denoting $\exists t < \tau$. We can classify the cases into two cases: $t < \tau$ appears on some contractions or not. The most stunning observation is that all contributions $\exists t < \tau$ at large and equal fictitious times are zero within correlation functions.

For the first case where connection arises between external leg $k\,(\tau\in[0, t])$ and internal line $p\,(\tau_{p}\in[0, t_{p}])$ with $t < t_{p}$, now the external fictitious time range $t$ is imposed on $t_2$ in (\ref{eq23}) while the internal one $t_{p}$ is imposed on $t_1$. So now there is a common term $\exp(-k^2 t_{p})$ which acts like a doubly-mentioned momentum while $\pm\exp(\pm k^2 t)$ can be extracted out of the whole integrals. If we stick to 3\textsuperscript{rd} rule in the stochastic Feynman rules, we recognize that the external leg $k$ can be explained by (leaving the 4\textsuperscript{th} rule) 
\begin{equation}
\label{eq41}
\text{(Not doubly-mentioned external momentum $k$) : }\exp(2k^{2}t) - 1 = \exp[(k^2 + k^2)t] - \exp[(k^2 - k^2)t].
\end{equation}
We interpret this as we put a check mark on the $k$-leg which has the smallest fictitious time range $t$, not variable, and write down $\pm k^2$ inside the exponents due to the contraction. So we recover our checkmark strategy not be only applied to internal fictitious-time variables $\tau$ but also external fictitious times $t$. For the ordinary case $\forall t > \tau$, we only have $\exp[(k^2 - k^2)t] = 1$ for all external legs following checkmark strategy so the strategy for connected lines still reasonable. Hence all propagators including external ones must be doubly checked at the end. Also, because $k$ and $p$ are checked once the other unconnected lines which have branching line $k = -p$ contain fixed $-p^2$ inside their corresponding exponents when we put check marks on them. Furthermore, we see that the 6\textsuperscript{th} rule in \hyperlink{subsubsection.3.2.2}{Section 3.2.2} now have changed integral ranges $\tau_{i}\in[t, \tau_{i - 1}]$ for unconnected lines with $t_{p} \leq \tau_{i}$. So the temporal integral range for the lastly unconnected external leg, saying $k_1 (\tau_{1}\in[0, t_{1}])$ in 5\textsuperscript{th} rule, is changed to $\tau_{1}\in[t, t_{1}]$ which implies that the contributions $\exists t < \tau$ at large and equal fictitious times give zero. In point of mapping, the terms $-1$ for $k_{i}$'s are replaced by exponential terms which share the same exponents with coexisting exponential terms but just differ by fictitious time factor, $t$; saying 
\begin{equation}
\label{eq42}
\exp[(k_i^2 + \mathcal{K}_i)t_1] - \exp[(k_i^2 + \mathcal{K}_i)t]
\end{equation}
where $\mathcal{K}_{i}$ is a linear set of squared momenta belonging to $k_i$ following the fictitious-time ordering. Similarly, for the connected line which has $t < \tau$, there is an extra term inside the second exponent in the curly bracket in (\ref{eq23}), $2k^2 t$, because we have to cut the temporal limit starting from $t$. Thus, for the case $\exists t < \tau$, we have to attach the additional term for not doubly-mentioned momenta of branched lines so our previous scheme for the branched lines $\pm\exp(\pm p^{2}\tau_{p})$ each is changed to $\exp(p^2 \tau_p)$ and $-\exp(-p^2 \tau_p)\exp(2p^2 t)$. We see that we back to (\ref{eq36}) when $t\,\rightarrow\,0$. We interpret this as the cutting point of initial temporal boundary gets back to origin.\footnote{Such a limit, however, gives zero amplitude due to (\ref{eq41}).} If there are more than two external legs, then it's convenient to divide the fictitious-time ordering into intervals between external fictitious time ranges; saying $t_1 > \cdots > t_2$, $t_2 > \cdots > t_3$, $\cdots$. Then we apply our scheme to the each sections separately by noticing lower temporal boundary of the each internal variables. 

The $t < \tau$ conditions can also be acquired within a given fictitious-time orderings keeping $t > \tau$ on all connections. When we consider a case where two internal lines $p_{1}\,(\tau_{p1}\in[0, \tau_{1}])$ and $p_{2}\,(\tau_{p2}\in[0, \tau_{2}])$ are connected where each of them are branched out from external legs $k_{1}\,(\tau_{1}\in[0, t_{1}])$ and $k_{2}\,(\tau_{2}\in[0, t_{2}])$. Supposing $\tau_{1} > \tau_{2}$ from the contraction, we can impose $t_{1} > \tau_{1} > t_{2} > \tau_{2}$. The whole rule in above paragraph is identically applied except that (\ref{eq41}) now becomes $1$. According to the checkmark strategy, after we once put a mark on $k_2$ first then this momentum will be marked again at $t_2$ so we generalize
\begin{equation}
\label{eq43}
\text{(Doubly-mentioned external momentum $k$) : }1
\end{equation}
which agrees with the previous statement $\forall t > \tau$. Through (\ref{eq41}), (\ref{eq43}), and other stochastic Feynman rules in \hyperlink{subsubsection.3.2.2}{Section 3.2.2} we rediscover (\ref{eq6}). Also we notice that after we finish the checkmark strategy all lines including external ones should be doubly marked.
%-----------------------------------------------------------------------------------------------------%

%--------------------------Generalized Stochastic Feynman Rules at Finite Fictitious Times--------------------------%
\subsubsection{Generalized Stochastic Feynman Rules at Finite Fictitious Times}
From the last section, we develop generalized stochastic Feynman rules at finite fictitious times for any given fictitious-time orderings, including external fictitious times, as followings: 
\begin{enumerate}
\item[*] The 1\textsuperscript{st}-4\textsuperscript{th} rules equally hold, i.e., (\ref{eq31}), (\ref{eq32}), (\ref{eq33}), and (\ref{eq34}).
\item[5.]  For unconnected propagators $k_1\,(0 \leq \tau_1 \leq t_1)$, $\cdots$, $k_N\,(0 \leq \tau_N \leq t_N)$ with a given fictitious-time ordering section $\cdots > t^{\prime} > \tau_1 > \cdots > \tau_N > t > \cdots$ where $t$ and $t^{\prime}$ are external fictitious time ranges, in order of which has a smaller fictitious-time variable, impose\footnote{The last product directly represents an adjustment of the amplitude for the system $\exists{t}<\tau$.}
\begin{equation}
\label{eq44}
\begin{aligned}
&\int_{t}^{\tau_{i - 1}} d\tau_i \Bigg\{\sum_{\xi_{i, 1},\,\cdots\,,\,\xi_{i, m}\,=\,\pm1}\left(\prod_{r = 1}^{m}\xi_{i, r}\right)\exp\Bigg[\Bigg(k_i^2 + \sum_{j = 1}^{m}\,\left(\xi_{i, j}\,p_{i, j}^{2}\right) - \sum_{k = m + 1}^{n - 1}\,\left(q_{i, k}^2\right)\Bigg)\tau_{i}\Bigg]\\
&\qquad\qquad\qquad\qquad\qquad\qquad\qquad\qquad\qquad\qquad\qquad\qquad\qquad\qquad\qquad\times\prod_{l = 1}^{m}\exp\left[\left(1 - \xi_{i, l}\right)p_{i, l}^{2}t\right]\Bigg\}
\end{aligned}
\end{equation}
where $p_{i, 1}, \cdots, p_{i, m}$ are not doubly-mentioned internal lines while $q_{i, m + 1}, \cdots, q_{i, n - 1}$ are doubly-mentioned internal lines, both branched from $k_{i}$. Here the last integral variable runs between those two external fictitious times $\tau_{1} \in [t, t^{\prime}]$.$\quad\rightarrow\quad$Generalization of (\ref{eq35}) and (\ref{eq36}).
\item[6.] Following the given fictitious-time ordering, for external legs, apply the checkmark strategy also on $t_{\text{ext}}$'s not for the legs and their branching lines but the legs and their connected momenta, i.e., (\ref{eq41}) and (\ref{eq43}).
\item[*] The 7\textsuperscript{th}-9\textsuperscript{th} rules equally hold, i.e., (\ref{eq37}), (\ref{eq38}), and (\ref{eq39}).
\end{enumerate}
We recognize that $(t, t^{\prime}) > \tau$'s on all connections and letting $t\,\rightarrow\,0$ return \hyperlink{subsubsection.3.2.2}{Section 3.2.2} while all possible orderings $\exists t < \tau$ give zero contribution at equal and large fictitious time because of the last integral $\tau_{1}\in[t, t^{\prime}]$.
%--------------------------------------------------------------------------------------------------------%

%--------------------------Stochastic Feynman Rules at Large Fictitious Times--------------------------%
\subsection{Stochastic Feynman Rules at Large Fictitious Times}
In this section, we make an establishment of the stochastic Feynman rules at large fictitious times; equilibrium limits. Our overall procedure and results are very similar with what Damgaard and Huffel did in Ref.~\cite{damgaard1987stochastic} but we review this and reinterpret in terms of doubly-mentioned momenta on fictitious-time ordering diagrams. We keep in mind our previous result that all contributions $\exists t < \tau$ are negligible at large fictitious times so we explore them only $\forall t > \tau$. We first start from the stochastic Feynman rules at finite fictitious times. It's crucial to figure out doubly-mentioned lines on fictitious-time ordering diagrams. We adopt a set notation to deliver this observation then we suggest a way to write down full propagators at the equilibrium from the fictitious-time ordering diagrams.
%-----------------------------------------------------------------------------------------------%

%------------------------Strategies to Write Down Propagators at Large Fictitious Times------------------------%
\subsubsection{Strategies to Write Down Propagators at Large Fictitious Times}
To deal with a given stochastic diagram of a specific type at large fictitious times, we observe what happens when we let the system reach equilibrium. From the 3\textsuperscript{rd} rule in the stochastic Feynman rules at finite fictitious times, we find that all exponential terms have at least one external momentum term in their exponents. However, when we expand the all exponential terms in evaluating correlation functions, we find that, if there is at least one undetermined internal momentum inside the exponent, then this term decays rapidly at the equilibrium. Therefore, our aim is focused on the exponential term which is comprised only of a sum of the squared external momenta, and this term will be canceled at equal fictitious times by the factors (\ref{eq33}). 

We start with an integral term with the shortest fictitious-time variable. When we first integrate this term according to the 5\textsuperscript{th} stochastic Feynman rule, we will get the terms where internal lines are included in the exponents. To remove these internal momentum terms, we have to consider the case where the exponential term for the following smallest fictitious-time variable in (\ref{eq36}) has the duplicated internal momentum terms overlapped with the internal momenta inside the just previous exponents. Since these doubly-mentioned momenta equally carry the “$-$" sign inside the exponent, to remove those momentum terms inside the exponents, the corresponding doubly-mentioned momenta inside the previous exponents should equally carry the “$+$" sign. Then these doubly-mentioned squared-momentum terms inside the contributing exponent will be canceled with each other within the advance integrals. In the same scheme, we conclude that, starting from the smallest fictitious-time variable, a surviving exponent at each integral step over a specific fictitious time has only the momenta which are not ever-duplicated with all momenta inside the previous exponents which have smaller fictitious-time variables than that of the specific one. In other words, when we conduct a specific temporal integral, we must observe which momenta are doubly mentioned compared to the momenta within the previously conducted integrals. Our work is to remove those duplicated momenta at each temporal integral. Then we find that we can finally get an exponential whose exponent is composed only of the external momenta (total sum of squared external momenta).

We investigate this logic from the perspective of the set notation. We assume that there are $N$ unconnected lines  $k_1 (0 \leq \tau_1 \leq t_1)$, $k_2 (0 \leq \tau_2 \leq t_2)$, $\cdots$, $k_N (0 \leq \tau_N \leq t_N)$ with a given time ordering $\forall t_{\text{ext}} > \tau_1 > \tau_2 > \cdots > \tau_N > 0$. We also define $U_i$ as a set of momenta $k$'s including $k_i$ and its branching lines $p_i$ (that means, the lines around the vertex). Starting from the propagator with its momentum $k_N$ which has the smallest fictitious-time variable $\tau_N \in [0 , t_N]$, the contributable integral term is 
\begin{equation}
\label{eq45}
\int_{0}^{\tau_{N - 1}}d\tau_N \exp\left[\left(\Sigma_{U_N} k^2\right)\tau_N\right].
\end{equation}
This integral gives us the first propagator factor, 
\begin{equation}
\label{eq46}
\frac{1}{\sum_{U_N} k^2}.
\end{equation}
This propagator term stands for the line with its momentum $k_N$. Now we have to consider the next momentum $k_{N-1}$ which has the next smallest fictitious-time variable $\tau_{N-1}$. We can represent the sum of $k^2$ separately by the momenta which are duplicated or not. We denote this as 
\begin{equation}
\label{eq47}
\int_{0}^{\tau_{N-2}} d\tau_{N-1} \exp\left[\left(\Sigma_{(U_{N-1} - \,U_{N})}k^2 - \Sigma_{(U_{N-1}\,\cap\,U_{N})}k^2\right)\tau_{N-1}\right]\int_{0}^{\tau_{N - 1}}d\tau_N \exp\left[\left(\Sigma_{U_N} k^2\right)\tau_N\right].
\end{equation}
Up to this calculation, this gives an additional propagator term for $k_{N-1}$,
\begin{equation}
\label{eq48}
\frac{1}{\sum_{(U_{N-1}\,\cup\,U_{N}) - (U_{N-1}\,\cap\,U_{N})} k^2}.
\end{equation}
For the next integral over $\tau_{N-2}$ with corresponding momentum $k_{N-2}$, the contributable integral is
\begin{equation}
\label{eq49}
\begin{split}
&\int_{0}^{\tau_{N-3}}d\tau_{N-2}\exp\left[\left(\Sigma_{(U_{N-2} - \,U_{N - 1} - \,U_{N})}k^2 - \Sigma_{(U_{N-2}\,\cap\,U_{N-1})}k^2 - \Sigma_{(U_{N-2}\,\cap\,U_{N})}k^2\right)\tau_{N-2}\right]\\
&\quad\times\int_{0}^{\tau_{N-2}} d\tau_{N-1} \exp\left[\left(\Sigma_{(U_{N-1} - \,U_{N})}k^2 - \Sigma_{(U_{N-1}\,\cap\,U_{N})}k^2\right)\tau_{N-1}\right]\int_{0}^{\tau_{N - 1}}d\tau_N \exp\left[\left(\Sigma_{U_N} k^2\right)\tau_N\right]. 
\end{split}
\end{equation}
In general, we have
\begin{equation}
\label{eq50}
\text{(Any intersection between more than two $U_i$'s)} = \varnothing
\end{equation}
because the connections occur only when two lines are met not more than two of them. Up to this point, this formula will give an additional propagator term for $k_{N-2}$ as 
\begin{equation}
\label{eq51}
\frac{1}{\sum_{(U_{N-2}\,\cup\,U_{N-1}\,\cup\,U_{N}) - (U_{N-2}\,\cap\,U_{N-1}) - (U_{N-2}\,\cap\,U_{N}) - (U_{N-1}\,\cap\,U_{N})}k^2}.
\end{equation}
By keeping on this work, we find that the propagator term for $k_i$ is, in general, 
\begin{equation}
\label{eq52}
\frac{1}{\sum_{(U_{i}\,\cup\,\cdots\,\cup\,U_{N}) - (U_{i}\,\cap\,U_{i - 1}) - \,\cdots\, - (U_{i}\,\cap\,U_{N}) - \,\cdots\, - (U_{N - 1}\,\cap\,U_{N})}k^2}.
\end{equation}
The propagator term should be terminated in terms of
\begin{equation}
\label{eq53}
\frac{1}{\small\sum_{\text{ext}}k_{\text{ext}}^2}
\end{equation}
where $k_{\text{ext}}$ are external momenta. Therefore, the contributing exponential term at the equilibrium would be 
\begin{equation}
\label{eq54}
\exp\left[\left(\Sigma_{\text{ext}}k_{\text{ext}}^2\right)t_1\right]
\end{equation}
and this exponential will be canceled with the foremost factor, 
\begin{equation}
\label{eq55}
\prod_{\text{ext}}\exp\left[-\left(k_{\text{ext}}^2\right)t_{\text{ext}}\right],
\end{equation} 
at equal and large fictitious times. In summary, when we let $U_{\text{c}}$ as a set of momenta of propagators where connections occur, with unconnected lines with momenta $k_1 (0 \leq \tau_1 \leq t_1)$, $\cdots$, $k_N (0 \leq \tau_N \leq t_N)$, then the full propagator terms within correlation functions with a given fictitious-time ordering $\forall t_{\text{ext}} > \tau_1 > \tau_2 > \cdots > \tau_N > 0$ are 
\begin{equation}
\label{eq56}
\langle \cdots \rangle_{\text{Eq}} \sim \left(\prod_{U_{\text{c}}}\frac{1}{k^2}\right)\left\{\prod_{i = 1}^{N}\frac{1}{\sum_{\left(\bigcup_{j = i}^{N}U_{j}\right) - \sum_{i \leq k < l}^{N}(U_{k}\,\cap\,U_{l})}k_{i}^{2}}\right\}.
\end{equation}
We suggest a diagrammatic representation for the propagator terms in (\ref{eq55}) as 
\begin{equation}
\label{eq57}
\langle \cdots \rangle_{\text{Eq}} \sim \left(\prod_{U_{\text{c}}}\frac{1}{k^2}\right)\times\{\frac{1}{\includegraphics[scale=0.4]{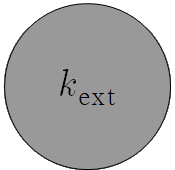}}\,\cdots\,\frac{1}{\includegraphics[scale=0.4]{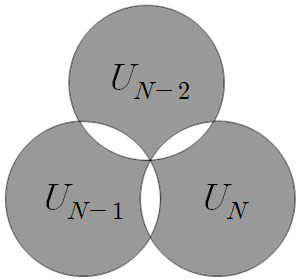}}\frac{1}{\includegraphics[scale=0.4]{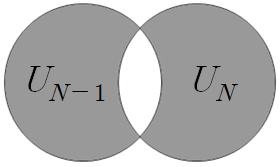}}\frac{1}{\includegraphics[scale=0.4]{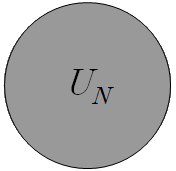}}\}.
\end{equation}
Each denominator in (\ref{eq57}) represents the sum of squared momenta, and these momenta are elements of the corresponding union (dark gray in each denominator). We denote our method in this paragraph as an \textbf{union strategy}. 

We can find all propagator terms from the fictitious-time ordering diagrams directly. We can first write down propagators for the connected lines by (\ref{eq34}). After then, starting from the unconnected propagator which has the smallest fictitious-time variable, we can write down a propagator for the momentum where its denominator is the sum of all squared momenta around the vertex of the momentum. In this step, it's convenient to put check marks on the mentioned momenta. Then when we are willing to write down the next propagator, we keep this strategy and combine the squared momenta with that in the previous propagator, but now we shouldn't count the already mentioned momenta. To figure out these doubly-mentioned momenta, we also put check marks on the momenta in this step, and we do not write down the momenta in the denominator if the momenta are doubly checked on the fictitious-time ordering diagram. The lastly remaining unconnected line can be described by a propagator whose denominator is composed of a sum of squared momenta of all external lines. We note that when we put check marks for $t$'s they just give $1$ for the external legs analogous to (\ref{eq43}).

We examine this from $\langle\phi(k_1)\phi(k_2)\rangle_{\text{NLO, Eq}}^{(1)}$ suggested in \autoref{Figure 2a} with its fictitious-time ordering diagram, \autoref{Figure 4}. We already showed that the propagator terms where connections occur are written as (\ref{eq25}). After we choose $\tau_1 > \tau_2$, the smallest fictitious-time variable $\tau_2$ stands for the unconnected (external) line with its momentum $k_2$ which has branching lines with momenta $\vert p_2 \vert = \vert p_1 \vert = p$ and $\vert q_2 \vert = \vert q_1 \vert = q$ so we put check marks on $k_2$, $p_2$, $p_1$, $q_2$, and $q_1$. Up to this point, we have 
\begin{equation}
\label{eq58}
\langle\phi(k_1)\phi(k_2)\rangle_{\text{NLO, Eq}}^{(1)} \sim \frac{1}{p^2}\frac{1}{q^2}\frac{1}{k_2^2 + p^2 + q^2}.
\end{equation}
For the last remaining unconnected line with its momentum $k_1$, which has the next smallest fictitious-time variable $\tau_1$, its branching lines have momenta $\vert p_1 \vert = \vert p_2 \vert = p$ and $\vert q_1 \vert = \vert q_2 \vert = q$ but they are already checked in the previous checkmark step. Hence, when we consider a union between the momenta $k_1$ and $k_2$ and their branching momenta, we shouldn't take account of the internal lines with momenta $p$ and $q$. Thus the union set has its elements of $k_1$ and $k_2$ which are both external lines so we write 
\begin{equation}
\label{eq59}
\langle\phi(k_1)\phi(k_2)\rangle_{\text{NLO, Eq}}^{(1)} \sim \frac{1}{p^2}\frac{1}{q^2}\frac{1}{k_2^2 + p^2 + q^2}\frac{1}{k_1^2 + k_2^2}.
\end{equation}
This result can be found in (\ref{eq13}). 
%-------------------------------------------------------------------------------------------------%

%--------------------------Generalized Stochastic Feynman Rules at Large Fictitious Times--------------------------%
\subsubsection{Generalized Stochastic Feynman Rules at Large Fictitious Times}
We explored the ways how to write down propagator terms in a given correlation function with a specific fictitious-time ordering. We suggest exact ways to solve it for $(D + 1)$-dimensional Euclidean $\phi^n$ theory. Other rules except for the rules describing unconnected lines are adopted from \hyperlink{subsubsection.3.2.2}{Section 3.2.2}: 
\begin{enumerate}
\item[*] The 1\textsuperscript{st}-4\textsuperscript{th} rules equally hold, i.e., (\ref{eq31}), (\ref{eq32}), (\ref{eq33}), and (\ref{eq34}).
\item[5.]  Among the unconnected propagators $k_1\,(0 \leq \tau_1 \leq t_1)$, $\cdots$, $k_N\,(0 \leq \tau_N \leq t_N)$ with a given fictitious-time ordering $\forall t_{\text{ext}} > \tau_1 > \cdots > \tau_N > 0$, pick the momentum $k_N$ which has the smallest fictitious-time variable $\tau_{\text{min}} = \tau_N$ as a starting point. If we define $U_i$ as a set of momenta $k$'s including $k_i$ and its branching momenta $p_i$, then, for the each unconnected line $k_i$, impose
\begin{equation}
\label{eq60}
\frac{1}{\sum_{\left(\bigcup_{j = i}^{N}U_{j}\right) - \sum_{i \leq k < l}^{N}(U_{k}\,\cap\,U_{l})}k_{i}^{2}}.
\end{equation}
Do this work up to the last unconnected line $k_1$ which gives
\begin{equation}
\label{eq61}
\frac{1}{\small\sum_{\text{ext}}k_{\text{ext}}^2}. 
\end{equation}
\item[6.] Multiply
\begin{equation}
\label{eq62}
\exp\left[\left(\Sigma_{\text{ext}}k_{\text{ext}}^2\right)t_{1}\right].
\end{equation}
\item[*] The 7\textsuperscript{th}-9\textsuperscript{th} rules equally hold, i.e., (\ref{eq37}), (\ref{eq38}), and (\ref{eq39}).
\end{enumerate}
At equal fictitious times (\ref{eq62}) canceled with (\ref{eq33}).
%------------------------------------------------------------------------------------------------%

%--------------Examination of One-Loop Structures in Two-Point Correlation Functions of $\phi^4$ Theory--------------%
\subsubsection{Examination of One-Loop Structures in $\phi^4$ Theory}
Following the stochastic Feynman rules suggested in \hyperlink{subsubsection.3.3.2}{Section 3.3.2}, we find that we have to take into account a strict consideration for one-loop structures in $\phi^4$ theory.

\begin{figure}
\centering
\includegraphics[width = 2.2in]{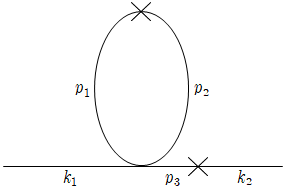}
\caption{\centering{Possible connections which constitute an one-loop contribution of two-point correlation functions in $\phi^4$ theory.}}
\label{Figure 5}
\end{figure}

We assume that an unconnected line with momentum $k_1\,(\tau_1 \in [0 , t_1])$ has its branching lines with momenta $p_1\,(\tau_{p1} \in [0 , \tau_1])$, $p_2\,(\tau_{p2} \in [0 , \tau_1])$, and $p_3\,(\tau_{p3} \in [0 , \tau_1])$ in $\phi^4$ theory with the coupling constant $\lambda$. We connect the first two of them so that they constitute an one-loop structure (there is one interaction with an order of $\lambda$) while the last branched line is connected with another external leg, $k_2\,(\tau_2 \in [0 , t_2])$ as we suggest in \autoref{Figure 5}.\footnote{To find an exact solution of the suggested two-point correlation function $\langle\phi(k_1, t_1)\phi(k_2, t_2)\rangle_{\eta}$ in $\phi^4$ theory, we have to sum total two kinds of topologically identical stochastic diagrams with their common symmetry factor $S = 3$, and apply the other stochastic Feynman rules at large fictitious times.} 
From those connections, saying $\vert p_1 \vert = \vert p_2 \vert \equiv p_{\text{loop}}$ and $\vert p_3 \vert = \vert k_2 \vert$, we first write down their corresponding propagators as 
\begin{equation}
\label{eq63}
\frac{1}{k_2^2}\frac{1}{p_{\text{loop}}^{2}}.
\end{equation}
For the (solely) unconnected line $k_1$, it seems to be imposed in the form of 
\begin{equation}
\label{eq64}
\frac{1}{k_1^2 + k_2^2 + 2p_{\text{loop}}^{2}}
\end{equation}
since the unconnected line is what we're first willing to discuss for unconnected lines so there is any doubly-mentioned momentum at all, but this gives a wrong answer. We can recognize this error because the propagator is not composed of a total sum of all squared external momenta, $k_1^2 + k_2^2$. When we consider the fictitious-time orderings embedded in the correlation function, since there is only one unconnected line, the fictitious-time orderings are trivially determined (the shortest fictitious time also be the last fictitious-time variable in the integral for the momentum $k_1$). Then we have one integral for the unconnected line with $k_1$ like (\ref{eq29}); that is 
\begin{equation}
\label{eq65}
\int_{0}^{t_1}d\tau_1 \left\{\sum_{\xi_{1},\,\xi_{2},\,\xi_{3}\,=\,\pm1}\xi_{1}\,\xi_{2}\,\xi_{3}\exp\left[\left(k_1^2 + \xi_{1}\,p_{\text{loop}}^{2} +\xi_{2}\,p_{\text{loop}}^{2} + \xi_{3}\,k_2^2\right)\tau_1\right]\right\}.
\end{equation}
Meanwhile, since the one-loop Feynman diagram is made of two equal momenta $p_{\text{loop}}$ which share equal fictitious time range $[0, \tau_1]$, then the common exponential factor $\exp(-p_{\text{loop}}^{2} \tau_1)$ appeared in the connection should be included in (\ref{eq65}); that means one $\pm p_{\text{loop}}^{2}$ should be fixed as $-p_{\text{loop}}^{2}$. Therefore, the correct formula for the $k_1$-leg with the loop structure at finite fictitious time is
\begin{equation}
\label{eq66}
\int_{0}^{t_1}d\tau_1 \left\{\sum_{\xi_{1},\,\xi_{2}\,=\,\pm1}\xi_{1}\,\xi_{2}\exp\left[\left(k_1^2 + \xi_{1}\,p_{\text{loop}}^{2} - p_{\text{loop}}^{2} + \xi_{2}\,k_2^2\right)\tau_1\right]\right\}
\end{equation}
which contains only 4 terms when we expand it. After doing this integral, the only contributable term at large fictitious time is the case where $\xi_1 = \xi_2 = +1$. Henceforth, at large fictitious times, by multiplying the connection propagators in (\ref{eq63}) and applying crossing symmetry between $k_1\,\leftrightarrow\,k_2$, we get
\begin{equation}
\label{eq67}
\frac{1}{k_2^2}\frac{1}{p_{\text{loop}}^{2}}\frac{1}{k_1^2 + k_2^2} + (k_{1}\,\leftrightarrow\,k_{2}) = \frac{1}{k_1^2}\frac{1}{k_2^2}\frac{1}{p_{\text{loop}}^2}.
\end{equation}
Again, this happens because the two momenta $p_1$ and $p_2$ connected with each other share the same momentum $p_{\text{loop}}$ and the same fictitious time range $(\tau_{p1}, \tau_{p2}) \in [0, \tau_1]$ where they are both branched out from one (external) line.\footnote{If one of them is branched from another line, then the fictitious-time ordering is to compare different fictitious times so the common exponential term in (\ref{eq23}) cannot be included in the temporal integral over the minimum fictitious-time variable anymore.} This directly states the one-loop Feynman diagram. Therefore, we conclude that 
\begin{remark}\label{remark1}
For loop structures where each loop has an order of $\lambda$, we do not include loop momenta in propagators for unconnected lines.
\end{remark}
We can derive this from our checkmark strategies. When we put a check mark on the unconnected momentum $k_1$, then its branching momenta, $\vert p_1 \vert = \vert p_2 \vert$ and $\vert p_3 \vert = \vert k_2 \vert$, should be marked but the one-loop momenta $p_1$ and $p_2$ are each doubly marked automatically due to the momentum conservation.\footnote{We see that $k_2$ is once checked by the momentum conservation when we put a mark on $p_3$.} Therefore, since two $p_{\text{loop}}$'s are doubly mentioned themselves simultaneously, we must not count those momenta at the large fictitious times. This is the reason why we impose a fixed $-p_{\text{loop}}^{2}$ term inside the exponents in (\ref{eq66}).
%------------------------------------------------------------------------------------------------------%

%--------------------------Examination of Propagators with Independent Fictitious-Time Variables--------------------------%
\subsubsection{Examination of Propagators with Independent Fictitious-Time Variables}
In the last section, our checkmark strategy was rigorously considered when there exist loop structures in $\phi^4$ theory. It appeared since the connection occurs between the lines whose momenta and fictitious time ranges are both identical. Meanwhile, we can also imagine a case where some unconnected internal lines share the same fictitious time ranges but have different momenta (as well as different fictitious-time variables). In this case, we find that the corresponding integrals are not necessarily conducted successively as (\ref{eq36}). 

We suppose that there are two unconnected lines with momenta $k_1\,(\tau_1 \in [0 , t])$ and $k_2\,(\tau_2 \in [0 , t])$, both branched from an unconnected line $k$ so that the two internal lines share the same fictitious time range.\footnote{Since the lines with momenta $k_1$ and $k_2$ are unconnected lines, the line with momentum $k$ where they are branched from is an unconnected line too.} When we apply the corresponding stochastic Feynman rules at the finite fictitious times, they have forms of\footnote{Here our ``removing doubly-mentioned momenta" strategy is already conducted for all previous unconnected lines.} 
\begin{equation}
\label{eq68}
\int_{0}^{t} d\tau_1 \left\{\exp[(k_1^2 + \mathcal{K}_1)\tau_1] \pm \cdots \right\}\int_{0}^{t} d\tau_2 \left\{\exp[(k_2^2 + \mathcal{K}_2)\tau_2] \pm \cdots \right\}.
\end{equation}
Because the two integrals above are independent calculations so our union strategy in (\ref{eq56}) at the large fictitious times is not necessarily applied between the lines with $k_1$ and $k_2$. However, the momenta $k_1$ and $k_2$ will be doubly marked when we put a check mark on $k$ in the next step so our union strategy is normally applied for the line $k$ but not for the lines $k_1$ and $k_2$, i.e., we should apply the strategy separately and independently to the lines $k_1$ and $k_2$. For instance, we suppose that a line with momentum $p_1$, branched from the line $k_1$, has its fictitious time interval as $\tau_{p1} \in [0 , \tau_1]$ while another line $p_2$ with its temporal interval $\tau_{p2} \in [0 , \tau_2]$ is branched from the line $k_2$. Then we apply the union strategy or checkmark strategy for $k_1$ by linking to $p_1$ while $k_2$ is linked to $p_2$ separately. If the branched lines $p_1$ and $p_2$ are unconnected ones too, then they will have their each union, saying $U_{p1}$ and $U_{p2}$, where the previously doubly-mentioned momenta are already offset. Therefore, the propagator terms for $k_1$ and $k_2$ are represented as 
\begin{equation}
\label{eq69}
\frac{1}{\Sigma_{(U_{k1}\,\cup\,U_{p1}) - (U_{k1}\,\cap\,U_{p1})}k^2}\frac{1}{\Sigma_{(U_{k2}\,\cup\,U_{p2}) - (U_{k2}\,\cap\,U_{p2})}k^2}.
\end{equation}
Here the set $U_{k1}$ and $U_{k2}$ each has elements of the line itself ($k_1$ and $k_2$) and its branching lines. After then, our union strategy is normally applied between the next unconnected $k$-line and previous unionized groups for $k_1$ and $k_2$. 

In general, this observation comes from where the two fictitious-time variables, $\tau_1$ and $\tau_2$, are independent, and the fictitious-time ordering between them is undetermined so that we cannot choose a specific fictitious-time ordering between $\tau_1$ and $\tau_2$. Recall that the 5\textsuperscript{th} rule in \hyperlink{subsubsection.3.3.2}{Section 3.3.2} stood for a continuously fully-given fictitious-time ordering, saying $\cdots < \tau < \tau^{\prime} < \tau^{\prime\prime} < \cdots$. However, if we have independent orderings as $\Theta(\tau^{\prime\prime} - \tau^{\prime})\Theta(\tau^{\prime\prime} - \tau)$ and there is no information about the orderings between $\tau$ and $\tau^{\prime}$, then we can conduct the integrals for the each time ordering independently. In the sense of the checkmark strategy, regardless of whether we pick the fictitious-time variable that has to be evaluated as $\tau$ or $\tau^{\prime}$, they give the same answer when we hold the separate union strategy as (\ref{eq69}); the sequences of the checkmark strategy $\cdots\,\rightarrow\,\tau\,\rightarrow\,\tau^{\prime}\,\rightarrow\,\cdots$ and $\cdots\,\rightarrow\,\tau^{\prime}\,\rightarrow\,\tau\,\rightarrow\,\cdots$ are equivalent.

In a more clear manner, we observe that regardless of whether we artificially choose fictitious-time ordering as either $\tau^{\prime\prime} > \tau^{\prime} > \tau$ or $\tau^{\prime\prime} > \tau > \tau^{\prime}$, the sum of these cases yields the same result as the one where the fictitious-time ordering between the independent variables $\tau$ and $\tau^{\prime}$ is undetermined, denoted as $\tau\xleftrightarrow[]{?}\tau^{\prime}$. For instance, for the previous example in (\ref{eq68}) and (\ref{eq69}),
\begin{equation}
\label{eq70}
\begin{aligned}
&\cdots\,\frac{1}{\Sigma_{(U_{k1}\,\cup\,U_{k2}\,\cup\,U_{p}) - (U_{k1}\,\cap\,U_{k2}) - (U_{k1}\,\cap\,U_{p}) - (U_{k2}\,\cap\,U_{p})}\,k^2}\frac{1}{\Sigma_{(U_{k2}\,\cup\,U_{p}) - (U_{k2}\,\cap\,U_{p})}\,k^2}\,\cdots\,(\tau_1 > \tau_2)\\
&\qquad + \cdots\,\frac{1}{\Sigma_{(U_{k2}\,\cup\,U_{k1}\,\cup\,U_{p}) - (U_{k2}\,\cap\,U_{k1}) - (U_{k2}\,\cap\,U_{p}) - (U_{k1}\,\cap\,U_{p})}\,k^2}\frac{1}{\Sigma_{(U_{k1}\,\cup\,U_{p}) - (U_{k1}\,\cap\,U_{p})}\,k^2}\,\cdots\,(\tau_2 > \tau_1)\\
&\qquad = \cdots\,\frac{1}{\Sigma_{(U_{k1}\,\cup\,U_{p1}) - (U_{k1}\,\cap\,U_{p1})}\,k^2}\frac{1}{\Sigma_{(U_{k2}\,\cup\,U_{p2}) - (U_{k2}\,\cap\,U_{p2})}\,k^2}\,\cdots\,(\tau_1\xleftrightarrow[]{?}\tau_2)
\end{aligned}
\end{equation}
where $U_{p}$ represents an union followed by an artificially fixed fictitious-time ordering up to right before $\tau_1$ or $\tau_2$ and the previously doubly-mentioned momenta are already offset within the union (to see a proof, refer \hyperlink{subsection.A.2}{Section A.2}). Thus, we conclude that 
\begin{remark}\label{remark2}
When dealing with unconnected lines where their corresponding fictitious-time variables are independent with each other, the union method is not necessarily applied between those lines. Instead, the method can be applied separately to the each line within its respective previously unionized group.
\end{remark}
%------------------------------------------------------------------------------------------------%

%----------------------------------------Conclusion----------------------------------------%
\section{Conclusion}
In this paper, we provided an overview of stochastic quantization for self-interacting scalar fields $\phi(x, t)$ with $\phi^3$ interaction. Utilizing perturbative solution of the Langevin equation, we empirically calculated two types of NLO of two-point correlation functions at both finite and large fictitious times. Then, for a simple discussion of correlation functions, we introduced a formalism for fictitious-time ordering diagrams which enabled us to figure out all possible fictitious-time orderings. Moreover, through the integration of this approach with our new method, checkmark strategy, we could directly write down contributions of specific stochastic diagram with its fictitious-time ordering, contingent upon whether the momentum lines are doubly mentioned. Based on this method, we established the stochastic Feynman rules at both finite ($\forall t > \tau$ and $\exists t < \tau$) and large fictitious times. We formulated the rules in both integral forms and fully expanded forms, utilizing the checkmark strategy. In the case of the rules $\exists t < \tau$, temporal cutoffs were applied to unconnected propagators, effectively serving as temporal initial conditions. One remarkable observation was that all contributions $\exists t < \tau$ approach zero at equilibrium. Our method of utilizing fictitious-time orderings provided more general descriptions to address calculating two specific situations; one-loop structures in $\phi^4$ theory and unconnected lines where the corresponding fictitious-time variables are independent. As a result, we generalized the stochastic Feynman rules at both finite and large fictitious times. We emphasize that our primary approach, directly utilizing fictitious-time orderings, is apparent in that it serves a unique understanding of correlation functions in stochastic quantization, especially for the non-equilibrium state.

We suggest some potential applications and open questions as follows:
\begin{itemize}
	\item We expect an analogous application to establishing generalized stochastic Feynman rules for fermions within a perturbative approach. Our fictitious-time ordering method for computing correlation functions can also serve as a convenient way for the same work for fermions.\footnote{We simply deduce this because it is inevitable to average temporal noise correlation functions between bosonic Green functions for fermions arising from the generalized Langevin equation. Furthermore, we can link our concept of connected and unconnected lines to bosonic and fermionic Green functions, holding the same forms as scalar fields with appropriate adjustments originating from a specifically chosen kernel. Detailed descriptions for the stochastic quantization for fermions can be found in Refs.~\cite{damgaard1987stochastic} and \cite{DAMGAARD198475}.} Also, one can practically apply the generalized rules to numerical simulations of a specific amplitude in the non-equilibrium correlation functions both $\forall t > \tau$ and $\exists t < \tau$.
	\item Together with the application for fermions above, we also anticipate that our generalized rules can be a reference for generalizing the rules to include gauge fields for gauge couplings (the gauge theories in stochastic quantization are dealt in Ref.~\cite{parisi1981perturbation} and \cite{damgaard1987stochastic}). Thus, applying our methodology of focusing on fictitious-time ordering will be advantageous in serving unified and simplified computation of correlation functions for fermionic and QED process and non-Abelian gauge theory at both finite and large fictitious times beyond scalar fields.
	\item In this study, our general methodologies and calculations were based on imaginary time and real scalar fields. However, we anticipate that our study for fictitious-time orderings can be extended to accomplish real-time stochastic Feynman rules at both equilibrium and non-equilibrium states. After this work, we will comprehend the real-time dynamics of each amplitude in non-equilibrium correlation functions both $\forall t > \tau$ and $\exists t < \tau$.
	\item So far, we have dealt with the case of weakly-coupled interactions with the corresponding perturbative approaches. Meanwhile, it's still challenging to come up with an idea of how we can further explore the fictitious-time orderings themselves for non-perturbative theories. Concentrating on lattice and quantum chromodynamics (QCD), this challenge should be addressed in future research, seeking another guideline of numerical simulations of a specific amplitude in a specific system at non-equilibrium.
\end{itemize}
%------------------------------------------------------------------------------------------%

%-----------------------------------------Acknowledgments-----------------------------------------%
\section*{Acknowledgments}
Moongul Byun would like to appreciate for a lot of support and guidance provided by Jae-Hyuk Oh.
%-------------------------------------------------------------------------------------------------%

%----------------------------------------------------Appendix----------------------------------------------------%
\appendix
\section{Appendix}
\subsection{Symmetry Factors in Stochastic Diagrams}
In \hyperlink{section.2}{Section 2} and \hyperlink{section.3}{3}, we adopted symmetry factors as multiplicative factors derived from Wick's theorem between the random fluctuations. We suggest a simple diagrammatic way to derive $S$ in this section.\cite{Musso:2006pt} When we calculated $\langle \eta_1 \cdots \eta_N \rangle_{\eta}$, we investigated ways that stochastic diagrams will not be changed under different connections between single stochastic diagrams, i.e., (\ref{eq3}). So the symmetry factor $S$ accounts for topologically identical (but have the same channel) stochastic diagrams at a given fictitious-time ordering. 
%----------------------------------------------------------------------------------------------------------------%

%-------------------Symmetry Factors of NLO in Two-Point Correlation Function in $\phi^3$ Theory-------------------%
\subsubsection*{Symmetry Factors of NLO in Two-Point Correlation Function in $\phi^3$ Theory}
In \hyperlink{section.2}{Section 2}, we derived that both $\langle \phi(k_1 , t_1) \phi(k_2 , t_2) \rangle_{\text{NLO}}^{(1)}$ and $\langle \phi(k_1 , t_1) \phi(k_2 , t_2) \rangle_{\text{NLO}}^{(2)}$ have the same symmetry factor, $S = 2$ (see (\ref{eq11}) and (\ref{eq17})). 
\begin{figure}
\centering
\includegraphics[width = 3.7in]{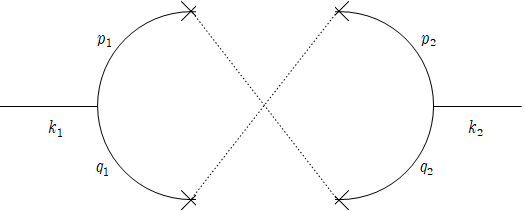}
\caption{\centering{Possible symmetric stochastic diagram of \autoref{Figure 2a}.}}
\label{Figure 6}
\end{figure}
Referring \autoref{Figure 6}, the first one can be obtained from the idea that chopping the diagram in \autoref{Figure 2a} into two single diagrams, rotating one of them, and re-connecting between the renewed diagrams.
%---------------------------------------------------------------------------------------------------------%

%----------------------Proof of Checkmark Strategy for Undetermined Fictitious-Time Orderings----------------------%
\subsection{Proof of Checkmark Strategy for Independent Fictitious-Time Variables}
In \hyperlink{subsubsection.3.3.4}{Section 3.3.4}, we concluded that there is no need to artificially choose fictitious-time orderings between independent fictitious-time variables. For a specific contribution within a specific correlation function, consider two unconnected lines with momenta $p (\tau_{p}\in[0, t_{p}])$ and $q (\tau_{q}\in[0, t_{q}])$ both originating from the same propagator. Suppose that the corresponding fictitious-time variables $\tau_{p}$ and $\tau_{q}$ are independent. Here we adopt a symbol $\mathcal{K}$ which was first introduced in (\ref{eq42}). When we let $\tau_{p} > \tau_{q}$, then the contribution within the correlation function is
\begin{equation}
\label{eq71}
\langle\cdots\rangle_{\text{Eq}} (\tau_{p} > \tau_{q}) = \cdots\frac{1}{p^2 + q^2 + \mathcal{K}_{p} + \mathcal{K}_{q} - \mathcal{K}_{p\,\cap\,q}}\frac{1}{q^2 + \mathcal{K}_{q}}\cdots
\end{equation}
where $\mathcal{K}_{p\,\cap\,q}$ represents the summation of squared doubly-mentioned momenta between $\mathcal{K}_{p}$ and $\mathcal{K}_{q}$. Conversely, if we let $\tau_{p} < \tau_{q}$, then 
\begin{equation}
\label{eq72}
\langle\cdots\rangle_{\text{Eq}} (\tau_{p} < \tau_{q}) = \cdots\frac{1}{q^2 + p^2 + \mathcal{K}_{q} + \mathcal{K}_{p} - \mathcal{K}_{p\,\cap\,q}}\frac{1}{p^2 + \mathcal{K}_{p}}\cdots.
\end{equation}
However, we note that $\mathcal{K}_{p\,\cap\,q} = 0$ for the lines with momenta $p$ and $q$ where their fictitious-time variables are independent. So, by summing (\ref{eq71}) and (\ref{eq72}), we obtain
\begin{equation}
\label{eq73}
\langle\cdots\rangle_{\text{Eq}} (\tau_{p}\xleftrightarrow[]{?}\tau_{q}) = \cdots\frac{1}{p^2 + \mathcal{K}_{p}}\frac{1}{q^2 + \mathcal{K}_{q}}\cdots
\end{equation}
as previously asserted in (\ref{eq70}).

%------------------Application of Stochastic Feynman Rules and Fictitious-Time Ordering Diagrams------------------%
\subsection{Application of Stochastic Feynman Rules and Fictitious-Time Ordering Diagrams}
Here we give applications of the stochastic Feynman rules at finite and large fictitious times with a fictitious-time ordering diagram where the two special cases \hyperref[remark1]{Remark 1} and \hyperref[remark2]{2} all appear in the last example.
%------------------------------------------------------------------------------------------------------%

%-------------------------Application to Two-Point Correlation Function in $\phi^3$ Theory-------------------------%
\subsubsection*{Application to Two-Point Correlation Function in $\phi^3$ Theory}
We apply our mapping strategy suggested in \hyperlink{subsubsection.3.2.2}{Section 3.2.2} to \autoref{Figure 2a}, $\langle\phi(k_1 , t_1)\phi(k_2 , t_2)\rangle_{\text{NLO}}^{(1)}$, for $(t_1, t_2) > \tau_1 > \tau_2$. Starting from the unconnected leg with $k_2$, referring \autoref{Figure 4}, we first have four terms 
\begin{equation}
\label{eq74}
\begin{aligned}
&\frac{1}{k_2^2 + p^2 + q^2}\left\{\exp\left[\left(k_2^2 + p^2 + q^2\right)t_1\right] - 1\right\}\,,\quad-\frac{1}{k_2^2 - p^2 + q^2}\left\{\exp\left[\left(k_2^2 - p^2 + q^2\right)t_1\right] - 1\right\}\,,\\
&\quad-\frac{1}{k_2^2 + p^2 - q^2}\left\{\exp\left[\left(k_2^2 + p^2 - q^2\right)t_1\right] - 1\right\}\,,\quad\frac{1}{k_2^2 - p^2 - q^2}\left\{\exp\left[\left(k_2^2 - p^2 - q^2\right)t_1\right] - 1\right\}.
\end{aligned}
\end{equation}
We need to take mapping from them onto 
\begin{equation}
\label{eq75}
\exp\left[\left(k_1^2 - p^2 - q^2\right)t_1\right] - 1.
\end{equation}
Then the full expansion of the two-point correlation function is, including connection terms, 
\begin{equation}
\label{eq76}
\begin{aligned}
\langle\phi(k_1 , t_1)\phi(k_2 , t_2)\rangle_{\text{NLO}}^{(1)}(\tau_1 > \tau_2)\sim\frac{1}{p^2}\frac{1}{q^2}\bigg(&\frac{1}{k_2^2 + p^2 + q^2}\frac{1}{k_1^2 + k_2^2}\left\{\exp\left[\left(k_1^2 + k_2^2\right)t_1\right] - 1\right\}\\
 - &\frac{1}{k_2^2 + p^2 + q^2}\frac{1}{k_1^2 - p^2 - q^2}\left\{\exp\left[\left(k_1^2 - p^2 - q^2\right)t_1\right] - 1\right\}\\
 - &\frac{1}{k_2^2 - p^2 + q^2}\frac{1}{k_1^2 + k_2^2 - 2p^2}\left\{\exp\left[\left(k_1^2 + k_2^2 - 2p^2\right)t_1\right] - 1\right\}\\
 + &\frac{1}{k_2^2 - p^2 + q^2}\frac{1}{k_1^2 - p^2 - q^2}\left\{\exp\left[\left(k_1^2 - p^2 - q^2\right)t_1\right] - 1\right\}\\
 - &\frac{1}{k_2^2 + p^2 - q^2}\frac{1}{k_1^2 + k_2^2 - 2q^2}\left\{\exp\left[\left(k_1^2 + k_2^2 - 2q^2\right)t_1\right] - 1\right\}\\
 + &\frac{1}{k_2^2 + p^2 - q^2}\frac{1}{k_1^2 - p^2 - q^2}\left\{\exp\left[\left(k_1^2 - p^2 - q^2\right)t_1\right] - 1\right\}\\
 + &\frac{1}{k_2^2 - p^2 - q^2}\frac{1}{k_1^2 + k_2^2 - 2p^2 - 2q^2}\left\{\exp\left[\left(k_1^2 + k_2^2 - 2p^2 - 2q^2\right)t_1\right] - 1\right\}\\
 - &\frac{1}{k_2^2 - p^2 - q^2}\frac{1}{k_1^2 - p^2 - q^2}\left\{\exp\left[\left(k_1^2 - p^2 - q^2\right)t_1\right] - 1\right\}\bigg).
\end{aligned}
\end{equation}
We see that only the first term contributes to the non-zero correlation function at equilibrium. 
%------------------------------------------------------------------------------------------------%

%----------------------Application to Two-Point Correlation Function in $\phi^4$ Theory----------------------%
\subsubsection*{Application to Two-Point Correlation Function in $\phi^4$ Theory}
\begin{figure}
	\centering
	\includegraphics[width = 3.8in]{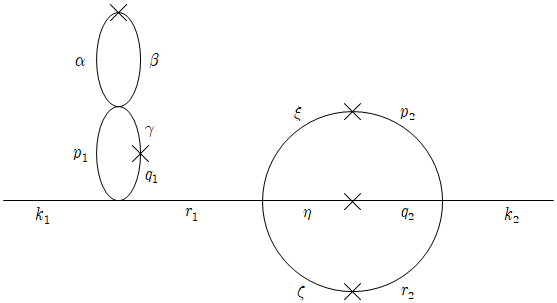}
	\caption{\centering{Stochastic diagram in two-point correlation function with $\phi^4$ interaction.}}
	\label{Figure 7}
\end{figure}

\begin{figure}
	\centering
	\includegraphics[width = 4.8in]{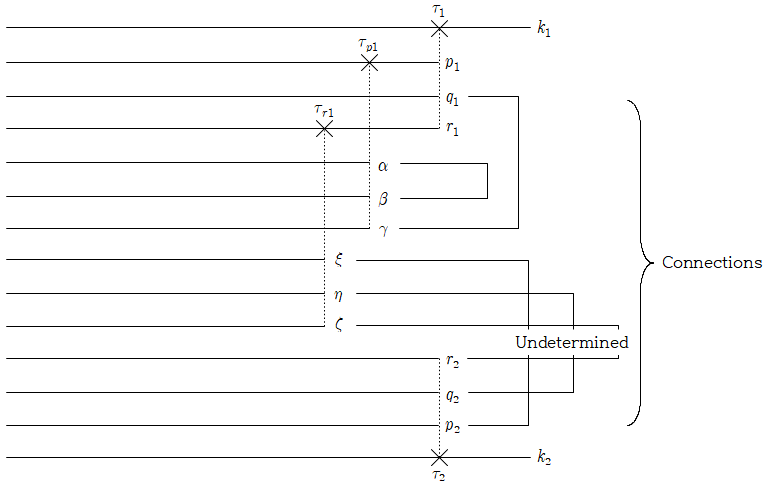}
	\caption{\centering{Fictitious-time ordering diagram which accounts for \autoref{Figure 7}.}}
	\label{Figure 8}
\end{figure}

We consider a stochastic diagram suggested in \autoref{Figure 7} which is not amputated.
Here the interaction is given by $\phi^4$ interaction. We define the propagators as $k_1\,(\tau_1 \in[0 , t_1])$, $k_2\,(\tau_2 \in [0 , t_2])$, $p_1\,(\tau_{p1} \in [0 , \tau_1])$, $q_1\,(\tau_{q1} \in [0 , \tau_1])$, $r_1\,(\tau_{r1} \in [0 , \tau_1])$, $p_2\,(\tau_{p2} \in [0 , \tau_2])$, $q_2\,(\tau_{q2} \in [0 , \tau_2])$, $r_2\,(\tau_{r2} \in [0 , \tau_2])$, $\alpha\,(\tau_{\alpha} \in [0 , \tau_{p1}])$, $\beta\,(\tau_{\beta} \in [0 , \tau_{p1}])$, $\gamma\,(\tau_{\gamma} \in [0 , \tau_{p1}])$, $\xi\,(\tau_{\xi} \in [0 , \tau_{r1}])$, $\eta\,(\tau_{\eta} \in [0 , \tau_{r1}])$, and $\zeta\,(\tau_{\zeta} \in [0 , \tau_{r1}])$. Then we can draw a corresponding fictitious-time ordering diagram as \autoref{Figure 8}.
From this fictitious-time ordering diagram, all possible fictitious-time orderings are given by 
\begin{equation}
\label{eq77}
\Theta(\tau_{1} - \tau_{p1})[\Theta(\tau_{r1} - \tau_{2}) + \Theta(\tau_{2} - \tau_{r1})]. 
\end{equation}
We choose $\Theta(\tau_{2} - \tau_{r1})$ with $(t_{1}, t_{2}) > \tau_1 > \tau_2 > (\tau_{r1}, \tau_{p1})$. For the connections, we directly impose 
\begin{equation}
\label{eq78}
\langle \phi(k_1)\phi(k_2) \rangle_{\text{Eq}} \sim \frac{1}{q_1^2}\frac{1}{\alpha^2}\frac{1}{p_2^2}\frac{1}{q_2^2}\frac{1}{r_2^2}. 
\end{equation}
Among the undetermined time ordering, we choose the smallest fictitious-time variable as $\tau_{r1}$ so we start from the unconnected line with its momentum $r_1$. We put a check mark on this momentum and its branching lines $\vert\xi\vert = \vert p_2 \vert$, $\vert\eta\vert = \vert q_2 \vert$, and $\vert\zeta\vert = \vert r_2 \vert$ so we find
\begin{equation}
\label{eq79}
\langle \phi(k_1)\phi(k_2) \rangle_{\text{Eq}} \sim \frac{1}{q_1^2}\frac{1}{\alpha^2}\frac{1}{p_2^2}\frac{1}{q_2^2}\frac{1}{r_2^2}\frac{1}{r_1^2 + p_2^2 + q_2^2 + r_2^2}. 
\end{equation}
When we consider the next unconnected line, $p_1$, however, we note that the $p_1$- and $r_1$-lines share the same fictitious time range, $(\tau_{p1}, \tau_{r1}) \in [0, \tau_1]$.\footnote{As we explained in \hyperlink{subsubsection.3.3.4}{Section 3.3.4} and \hyperlink{subsection.A.2}{A.2}, the same result is obtained regardless of whether we choose the minimum fictitious-time variable as $\tau_{p1}$ or $\tau_{r1}$.} Therefore, as we discussed in \hyperlink{subsubsection.3.3.4}{Section 3.3.4}, we don't take a union between the momentum groups for $p_1$ and $r_1$. So we put check marks on $p_1$ and its branching momenta $\vert\alpha\vert = \vert\beta\vert$, and $\vert\gamma\vert = \vert q_1 \vert$. But we also notice that the two lines $\vert\alpha\vert = \vert\beta\vert$ construct an one-loop diagram occurred at a spacetime so we do not count these momenta in the propagator for $p_1$. For these reasons, we write 
\begin{equation}
\label{eq80}
\langle \phi(k_1)\phi(k_2) \rangle_{\text{Eq}} \sim \frac{1}{q_1^2}\frac{1}{\alpha^2}\frac{1}{p_2^2}\frac{1}{q_2^2}\frac{1}{r_2^2}\frac{1}{r_1^2 + p_2^2 + q_2^2 + r_2^2}\frac{1}{p_1^2 + q_1^2}. 
\end{equation}
For the next unconnected external line $k_2$, now we have to take a union between the momentum groups for $k_2$, $p_1$, and $r_1$. Therefore, not counting the loop momentum, since the branching lines of $k_2$ are $p_2$, $q_2$, and $r_2$ which were all already once mentioned, we write 
\begin{equation}
\label{eq81}
\langle \phi(k_1)\phi(k_2) \rangle_{\text{Eq}} \sim \frac{1}{q_1^2}\frac{1}{\alpha^2}\frac{1}{p_2^2}\frac{1}{q_2^2}\frac{1}{r_2^2}\frac{1}{r_1^2 + p_2^2 + q_2^2 + r_2^2}\frac{1}{p_1^2 + q_1^2}\frac{1}{k_2^2 + p_1^2 + q_1^2 + r_1^2}. 
\end{equation}
After then, we apply our stochastic Feynman rules to the last unconnected, and not ever checked line, $k_1$. Since the branching lines of $k_1$-line were already once checked, we do not count these momenta. So we find 
\begin{equation}
\label{eq82}
\langle \phi(k_1)\phi(k_2) \rangle_{\text{Eq}} \sim \frac{1}{q_1^2}\frac{1}{\alpha^2}\frac{1}{p_2^2}\frac{1}{q_2^2}\frac{1}{r_2^2}\frac{1}{r_1^2 + p_2^2 + q_2^2 + r_2^2}\frac{1}{p_1^2 + q_1^2}\frac{1}{k_2^2 + p_1^2 + q_1^2 + r_1^2}\frac{1}{k_1^2 + k_2^2}. 
\end{equation}
We notice that the denominator in the last propagator (standing for $k_1$) exactly follows (\ref{eq61}). Furthermore, as we discussed in \hyperlink{subsubsection.3.3.4}{Section 3.3.4}, it's easy to prove 
\begin{equation}
\label{eq83}
\langle \phi(k_1)\phi(k_2) \rangle_{\text{Eq}}(\tau_1\xleftrightarrow[]{?}\tau_2) = \langle \phi(k_1)\phi(k_2) \rangle_{\text{Eq}}(\tau_1 > \tau_2) + \langle \phi(k_1)\phi(k_2) \rangle_{\text{Eq}}(\tau_1 < \tau_2).
\end{equation}
%---------------------------------------------------------------------------------------------------------%

%--------------------------------------------------References--------------------------------------------------%
\printbibliography
%\bibliographystyle{h-physrev.bst}
%\bibliography{ref}
%--------------------------------------------------------------------------------------------------------------%

\end{document}